\documentclass[sigconf,authorversion,nonacm]{acmart}

\usepackage{amsmath}

\usepackage{array}

\makeatletter
\newcolumntype{"}{@{\hskip\tabcolsep\vrule width 1pt\hskip\tabcolsep}}
\makeatother

\copyrightyear{2021} 
\acmYear{2021} 
\setcopyright{acmlicensed}\acmConference[iiWAS2021]{The 23rd International Conference on Information Integration and Web Intelligence}{November 29-December 1, 2021}{Linz, Austria}
\acmBooktitle{The 23rd International Conference on Information Integration and Web Intelligence (iiWAS2021), November 29-December 1, 2021, Linz, Austria}
\acmPrice{15.00}
\acmDOI{10.1145/3487664.3487673}
\acmISBN{978-1-4503-9556-4/21/11}

\setcopyright{none}
\begin{document}
	
	\title{RevASIDE: Assignment of Suitable Reviewer Sets for Publications from Fixed Candidate Pools}
	
	
	\author{Christin Katharina Kreutz}
	\email{kreutzch@uni-trier.de}
	\orcid{0000-0002-5075-7699}
	\affiliation{%
		\institution{Trier University}
		\city{Trier} 
		\country{Germany}
	}
	
	\author{Ralf Schenkel}
	\email{schenkel@uni-trier.de}
	\orcid{0000-0001-5379-5191}
	\affiliation{%
		\institution{Trier University}
		\city{Trier} 
		\country{Germany}
	}

	
	\begin{abstract}
		Scientific publishing heavily relies on the assessment of quality of submitted manuscripts by peer reviewers. Assigning a set of matching reviewers to a submission is a highly complex task which can be performed only by domain experts.
		We introduce RevASIDE, a reviewer recommendation system that assigns suitable sets of complementing reviewers from a predefined candidate pool without requiring manually defined reviewer profiles. Here, suitability includes not only reviewers’ expertise, but also their authority in the target domain, their diversity in their areas of expertise and experience, and their interest in the topics of the manuscript. 
		We present three new data sets for the expert search and reviewer set assignment tasks and compare the usefulness of simple text similarity methods to document embeddings for expert search. Furthermore, an quantitative evaluation demonstrates significantly better results in reviewer set assignment compared to baselines. A qualitative evaluation also shows their superior perceived quality.
	\end{abstract}

	\ccsdesc[500]{Information systems~Expert search}
	\ccsdesc[500]{Information systems~Recommender systems}
	
	\keywords{reviewer assignment, recommendation system, expertise modelling}

	\maketitle
	
	\section{Introduction}
	
	Peer review is a popular method of ensuring scientific standards for conferences and journals. It requires the assignment of suitable experts for each submission, which is often done manually~\cite{chughtai}. These reviewers then provide objective assessment of the manuscript and recommend accepting or rejecting the submission~\cite{ishag}.
	All of this has to be performed in a tight time frame~\cite{toronto}.
	
	The continuously increasing number of submissions as well as the high complexity of the task even for experienced chairs of program committees or journal editors calls for fully automatic methods of expert assignment. Furthermore, it is not sufficient to focus on the quality of single reviewers, but a \emph{good set of complementing reviewers} should be recommended for each manuscript. The \emph{reviewer assignment problem} tackles the task of retrieving sets of suitable reviewers for manuscripts submitted to a venue.
	
	%
	Even though the construction of sets of reviewers fitting submitted manuscripts has been studied frequently, most work focuses on construction of sets with the highest possible expertise but does not consider (m)any other aspects. Such other aspects could help  reduce reviewers' work load and increase comprehensiveness of reviews. Additionally, actual human evaluation of the sets and thus a reliable confirmation of results is generally not conducted.
	
	Numerous works~\cite{toronto,ishag,trendResInt2,DBLP:journals/pvldb/KouUMLLG15,robustModel,DBLP:conf/aciids/MaleszkaMKHMHV20,DBLP:journals/asc/YangLYCN20,mlClass} tackle the reviewer assignment problem in different ways, with slightly different definitions for the suitability of reviewers. While expertise of a reviewer with the topic of the manuscript~\cite{toronto,ishag,trendResInt2,DBLP:journals/pvldb/KouUMLLG15,robustModel,DBLP:journals/asc/YangLYCN20,mlClass} has been dominating in existing work, other features like authority~\cite{trendResInt2,robustModel}, research interest~\cite{trendResInt2} and diversity~\cite{ishag,robustModel,DBLP:conf/aciids/MaleszkaMKHMHV20} were considered in some existing work, but not in a holistic way. Additionally, these aspects were defined heterogeneously in present works.
	We incorporate the following five aspects into our definition of suitability of reviewer sets: \textit{expertise} of reviewers in general topics and methods of a submission, \textit{authority} of reviewers in the domain of the manuscript, \textit{diversity} in terms of reviewers differing in their areas of expertise, \textit{interest} of reviewers in the topics of the submission and diversity in terms of \textit{seniority} aspects of the reviewer set.
	
	In this work, we embark on finding the best reviewer sets for a submitted scientific paper from a predefined candidate pool in terms of these five aspects. 
	Unlike some existing work~\cite{sigmod,DBLP:conf/wise/PapagelisPN05}, we explicitly do not require the manual definition of keywords or bids on manuscripts from reviewer candidates.
	To achieve this, we make two important contributions: 
	\textit{1)} We propose and thoroughly evaluate RevASIDE, a new and completely automated technique for recommending sets of reviewers from a fixed set of candidates for single manuscripts. For this we introduce seniority as a completely new aspect and its combination with already established but redefined features.
	\textit{2)} We publish three different data sets suitable for expert search as well as reviewer set assignment.
	
	While we build on established expert retrieval methods to find reviewers with high expertise, our method is the first to incorporate all of the complementary factors authority, diversity, interest of candidates and seniority to solve the reviewer set assignment problem. 
	Our approach consists of two steps. Step 1 identifies topically relevant reviewers based on the similarity of their research direction to the manuscript, utilising expert search methods. Step 2 then assembles sets from these experts and determines the reviewer set that performs best in the five aspects. 
	To the best of our knowledge, this is the first work that utilises the expert search task as a preparatory step for the reviewer set assignment task.
	
	\section{Related Work}
	\label{rw}
	
	Retrieval-based approaches for scientific reviewer assignment treat the manuscript for which reviewers are searched as a query. They determine fitting reviewers based on different aspects, often under additional constraints. Such methods can be divided into ones recommending single reviewers for manuscripts, so-called expert search, and those tackling the assignment of whole reviewer sets.
	
	\subsection{Expert Search}
	Several papers target the recommendation of single reviewers for manuscripts which contrasts our goal of recommending reviewer sets. We identify and assemble the best fitting experts to a suitable set while the following works only handle the expert search task which disregards set effects.
	
	Numerous works pursue the expert search task as a matching problem between the query manuscript and expert profiles formed by their past publications. Some of them also consider more aspects than textual similarity: 
	MINARET~\cite{minaret} is a recommendation framework based on publications and affiliations of experts as well as expanded keywords for manuscripts. After an initial filtering step, it returns ranked list of reviewers. Candidates receive a score based on topical coverage, impact, recency, experience in reviewing and their familiarity with the target venue. 
	Chughtai et al.~\cite{chughtai} suggest ontology-based and topic-specific recommendation of single experts fitting a submission.
	Macdonald and Ounis~\cite{macdonald} propose twelve voting techniques to find suitable experts for query manuscripts. These techniques base on similarity of the reviewer candidates and the manuscripts. We use and extend their methods in Step 1 of our approach.
	
	Other works transform single expert finding into a classification problem: Yang et al.~\cite{colabIntell} base their approach on word-semantic relatedness via Wikipedia. Reviewers are ranked with respect to a manuscript by experience in the domain of the submission and their number of papers. Zhao et al.~\cite{Zhao} utilise word embeddings of keywords from author profiles and manuscripts to propose fitting reviewers. Similar to this approach we use embedding methods to abstract from words while searching for reviewer candidates. 
	
	\subsection{Reviewer Set Recommendation}
	
	Reviewer set recommendation can be observed for \textit{single papers} or \textit{multiple/all papers of a venue}. 
	The following approaches tackle reviewer set recommendation but consider different or fewer aspects compared to RevASIDE for estimating the quality of reviewer sets.
	
	%
	Ishag et al.~\cite{ishag} incorporate the $h$ index of reviewers, citation counts and paper diversity into their approach based on itemset mining. They return reviewer sets fitting a query manuscript and estimate the sets' impact. Contrasting their definition of diversity which uses the number of different affiliations of authors of a single paper, we define diversity as a measure between authors to estimate the actual topical differences in reviewer sets.
	Maleszka et al.~\cite{DBLP:conf/aciids/MaleszkaMKHMHV20} tackle the reviewer set assignment problem for one manuscript at a time by focusing on diversity aspects in expertise, the co-authorship graph and style of reviewers. They begin the set recommendation process with a single reviewer determined by another method. 
	Zhang et al.~\cite{mlClass} utilise a multi-label classifier for the construction of reviewer sets. The approach bases on predicted research labels for manuscripts and predicts reviewers with similar labels. Set-based effects are ignored which contrasts our approach.
	
	Works tackling the reviewer set recommendation for \textit{multiple papers} can be divided in ones relying on \textit{manual inputs} such as bidding by reviewers and \textit{fully automated} ones.
	Some of the papers incorporating \textit{manual inputs}, contrasting our fully automated method, are the following:
	The Toronto Paper Matching System (TPMS)~\cite{toronto} conducts automatic reviewer assignment for all manus\-cripts submitted to a conference by using either word count representation or LDA topics, but can also incorporate reviewers' bids on submissions. TPMS supports some constraints: papers must be reviewed by three reviewers, and reviewers are assigned not more than a given limit of papers. Reviewers for manuscripts are determined based on expertise extracted from their publications. 
	TPMS is applied, for example, by the SIGMOD research track~\cite{sigmod}, where reviewers upload a representative set of their publications. 
	%
	Papagelis et al.~\cite{DBLP:conf/wise/PapagelisPN05} present a system which incorporates reviewers' interests in terms of paper topics, their bids on papers, conflicts of interests and overall workload balance for the reviewer assignment task. It can either assign reviewer sets automatically if the bidding is completed or the PC chair can manually adjust the sets.
	
	The following works are \textit{fully automated} recommendation approaches intended to work with multiple manuscripts\footnote{Note that we currently refrain from this task as it would require an evaluation data set which includes all submissions to the venue, even the rejected ones and their authors. Such a data set does not exist currently to the best of our knowledge.}:
	Liu et al. \cite{robustModel} recommend $n$ reviewers for each manuscript which are dependent on each other. They model reviewers' expertise, authority and diversity as a graph which they traverse with random walk with restart. The number of co-authorships is modelled as authority which contrasts our definition of authority. 
	Kou et al.~\cite{DBLP:journals/pvldb/KouUMLLG15} introduce an assignment system for sets of $n$ reviewers which bases on the topic distributions of reviewers and the manuscripts computed with the Author-Topic Model. They define expertise of reviewer sets in certain topics as the maximum expertise for the topic found in the set; our definition of expertise deviates.
	Jin et al~\cite{trendResInt2} assume reviewers have a certain relevance in a topic which is determined by their publications and usage of the Author-Topic Model. Additionally, authority in form of citations and research interest of researchers are important factors. Here, the number of reviewers per paper and the maximum number of papers a reviewer is assigned to can be predefined. Amongst others we also observe these factors but define them differently. 
	Yang et al.~\cite{DBLP:journals/asc/YangLYCN20} utilise LDA to represent manuscripts as well as past publications of reviewer candidates. They then use a discrete optimisation model which focuses on expertise to assign reviewers to all manuscripts. Likewise, we also incorporate LDA in our approach but we additionally consider more aspects beyond expertise.

	\section{Aspects}
	\label{aspects}
	
	In our work, we assess the appropriateness of a reviewer set with respect to a submission based on the following seven aspects:
	
	\textbf{Aspect 1}
	Reviewers in a reviewer set cannot have \textit{conflicts of interests}: they can be neither authors of the submission nor prior co-authors of its authors~\cite{DBLP:conf/wise/PapagelisPN05}. This aspect aims at ensuring unbiased and objective candidates. While we (as well as others~\cite{DBLP:conf/wise/PapagelisPN05}) regard this aspect quite vigorously, less restrictive variants (e.g. disallowing co-authorships in the three years prior to the submission) are also feasible.
	
	\textbf{Aspect 2}
	Reviewers cannot be co-authors of any other reviewer in the set. Reviewers having \textit{disjoint publications} enforces a broader spectrum of different backgrounds. This could produce broader reviews~\cite{DBLP:conf/aciids/MaleszkaMKHMHV20} which is a desirable property in peer review~\cite{sigmod}. 
	
	\textbf{Aspect 3}
	Reviewers need to be \textit{experienced} in the area of the manuscript~\cite{DBLP:journals/pvldb/KouUMLLG15}. The topic of the paper should be relevant for them and fit their research profile. Not only the content but also the number of papers in the area of a submission contributes to our understanding of experience. This aspect ensures deep reviews, another desirable feature of assessments~\cite{sigmod}.
	
	\textbf{Aspect 4}
	Reviewers need to hold \textit{authority} in the research area of the submission. Reviews of the papers have to be credible, reviewers should be well recognised in the target domain~\cite{robustModel}. Authority can be assessed, for example, by an area-dependent $h$ index and citation counts of candidates.
	
	\textbf{Aspect 5}
	Reviewers need to be \textit{diverse} in their area of expertise. Typically, as many topics as possible of a submission should be assessed to create a comprehensive review~\cite{sigmod}. Reviewers that are proficient in different topics from each other support this goal as the candidates in a set have unique perspectives formed by their different experiences and backgrounds~\cite{DBLP:conf/aciids/MaleszkaMKHMHV20}.
	
	\textbf{Aspect 6}
	Reviewers need to be \textit{currently interested} in the topics of the manuscripts so they accept the reviewing request~\cite{trendResInt2} and are not asked to review topics they no longer work in. Scientific progress makes it impossible to be up to date in all areas they were formerly interested in. Thus, time-aware suggestion should weigh recent works of reviewers much higher than older publications.
	
	\textbf{Aspect 7}
	Reviewers of a manuscript should not solely consist of \textit{senior} researchers, but they need to be diverse with respect to the amount of their experience. Senior researchers provide vast reviewing experience and a global vision but they should be handled as a sparse resource as they are asked to review many submissions. Junior researchers are ambitious and resilient while not having that much experience. Usually, they are less frequently asked to review and more of an unexhausted resource.
	Reviewing load needs to be distributed between senior and junior researchers such that the lower load for senior researchers and incorporation of newer researchers benefits the overall quality of reviews. 
	Additionally, junior researchers could provide new and refreshing perspectives while the reviewing activity might also benefit their own development.
	Breaking up well-established reviewer constellations with new candidates could also avoid research cliques~\cite{toronto}.
	
	\section{Approach}
	\label{revaside}
	
\begin{figure}[t]
		\centering
		\includegraphics[width=0.475\textwidth]{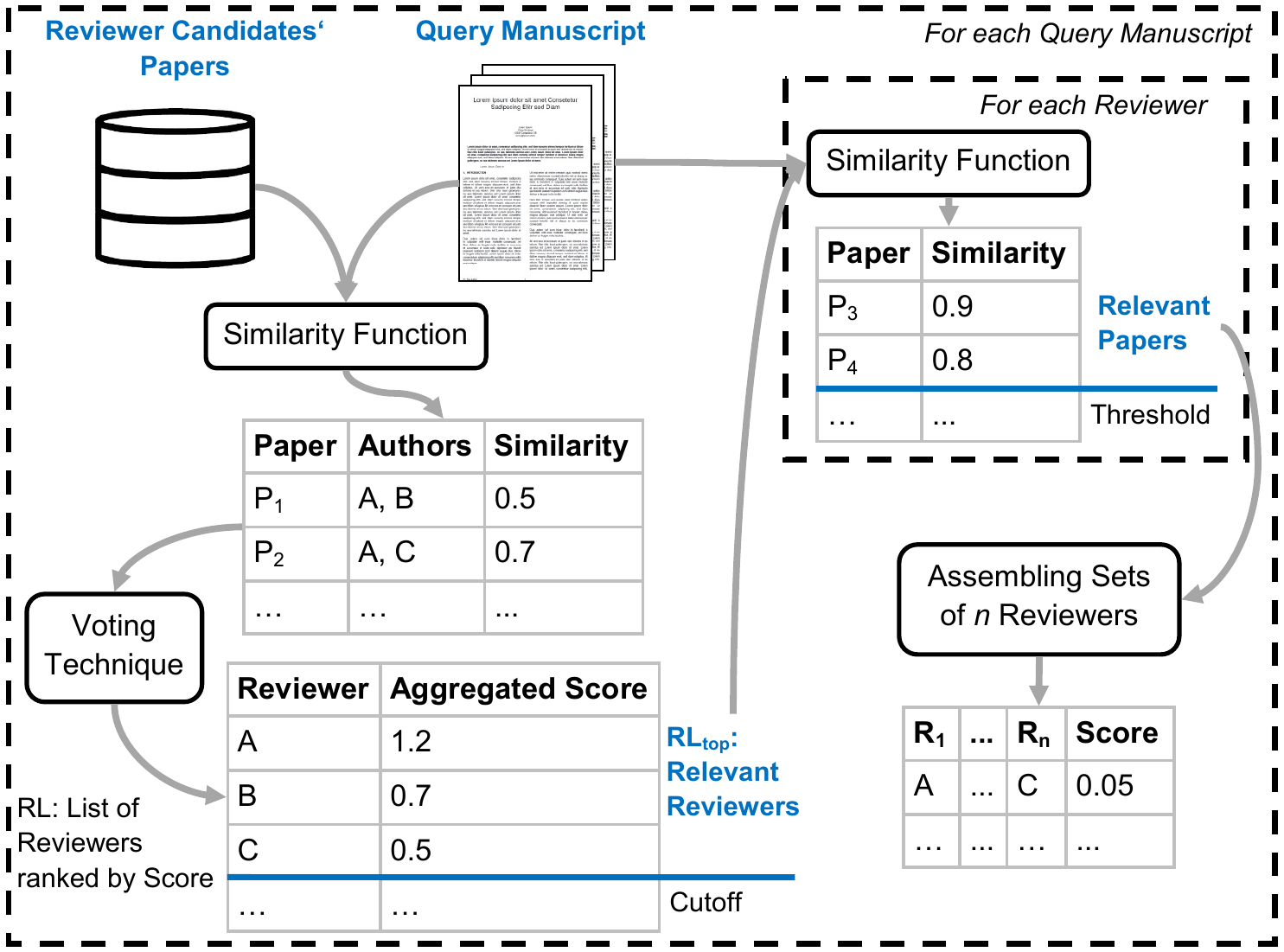}
		\caption{Schematic overview of our approach.
			The left part depicts the expert search task, the right part depicts the set of reviewers assignment task.
		}
		\label{fig:steps}
	\end{figure}

	RevASIDE is a system for assigning sets of \textbf{Rev}iewers utilising \textbf{A}uthority, \textbf{S}eniority, \textbf{I}nterest, \textbf{D}iversity and \textbf{E}xpertise of reviewers to find the most suitable reviewer set out of a fixed set of candidates, the reviewer candidate pool $RCP$, for a given manuscript $M$. Our approach is composed of two steps: in Step 1, suitable reviewers are identified from the pool of reviewer candidates; in Step 2, they are assembled to the most suitable set for the manuscript. Figure~\ref{fig:steps} depicts the schematic overview of our approach.
	
	\subsection{Step 1: Expert Search}
	
	Step 1 handles the left part of Figure~\ref{fig:steps}.
	We represent publications as tf-idf vectors or ones constructed with BERT~\cite{BERT} or Doc2Vec~\cite{d2v}, which allows to depict semantics of documents instead of single tokens. This enables capturing similarity of concepts of papers. 
	
	Let $M$ be the manuscript for which a reviewer set should be computed. We ignore any reviewers for which a conflict of interest with the authors of $M$ exists (Aspect 1). 
	For the remaining reviewers from the reviewer candidate pool $RCP$, let $P(R)$ be the set of publications written by reviewer $R$. The similarity between a publication $P$ and a manuscript $M$ is given by $sim(P, M)$; the utilised similarity measure can be changed between the two steps. In our experiments, we will use the cosine similarity of the corresponding vectors.
	We then sort $R$'s papers in descending order by their similarity to the manuscript $M$ and denote by $rank(P, R, M)$ the rank of a certain publication $P$ of reviewer $R$ in this order. Similarly, we sort all publications in the collection in descending order by their similarity to manuscript $M$ and denote by $rank(P, M)$ the rank of a publication $P$ in this order. 
	
	\begin{table}[t]
		\caption{Voting techniques VT and accompanying formulas for reviewer $R$ and manuscript $M$.}
		\label{tab:vt}
		\centering
		\begin{tabular}{c|l}
			VT & Formula\\ \hline
			$Votes_{\delta}$ & $\sum_{P \in P(R) \land sim(P, M) \ge \delta} 1$\\
			$SUM$ & $\sum_{P \in P(R)} sim(P, M)$\\
			$AVG$ & $\frac{\sum_{P \in P(R)} sim(P, M)}{|P(R)|}$\\
			$MNZ$ & $|P(R)| * \sum_{P \in P(R)} sim(P, M)$\\
			$SUM_{n}$ & $\sum_{P \in P(R) \land rank(P, R, M) \le n} sim(P, M)$\\
			$MIN$ & $min\big(\{sim(P, M) | P \in P(R)\}\big)$\\
			$MAX$ & $max\big(\{sim(P, M) | P \in P(R)\}\big)$\\
			$RR$ & $\sum_{P \in P(R)} \frac{1}{rank(P, M)}$\\
			$mRR$ & $\frac{1}{|P(R)|} \sum_{P \in P(R)} \frac{1}{rank(P, M)}$\\
			$BordaFuse$ & $\sum_{P \in P(R)} (|\bigcup_{R_i \in RCP} P(R_i)| - rank(P, M))$\\
			$exp^{SUM}$& $\sum_{P \in P(R)} e^{sim(P, M)}$\\
			$exp^{AVG}$ & $\frac{\sum_{P \in P(R)} e^{sim(P, M)}}{|P(R)|}$\\
			$exp^{MNZ}$ & $|P(R)| * \sum_{P \in P(R)} e^{sim(P, M)}$
		\end{tabular}
	\end{table}{}
	To obtain a ranked list $RL$ of reviewers, we apply a number of voting techniques (VTs) that score reviewer candidates with respect to a manuscript. These voting techniques base on the ones applied by Macdonald and Ounis~\cite{macdonald} for expert search. Table \ref{tab:vt} shows the exact formulas for the 13 voting techniques considered in our approach. Higher scores signal better fit of a reviewer to the given manuscript. $Votes_\delta$ computes the number of papers of a reviewer with a similarity to the query manuscript not smaller than a threshold $\delta$; note that the method was introduced without such a threshold in~\cite{macdonald}, which corresponds to $\delta=0$ in our definition. $SUM$ sums up the similarities of the papers of a reviewer with the query manuscript, $AVG$ uses this score and normalizes it by the total number of papers of the reviewer. $MNZ$ multiplies the $SUM$ score by the number of papers of the reviewer. $SUM_{n}$ sums the similarities of the $n$ papers of the reviewer most similar to the manuscript. $MIN$ returns the smallest similarity of the reviewer's paper with the manuscript, $MAX$ is defined analogously. $RR$ sums up the reciprocal ranks of the reviewer's papers in the ordered list of all papers. We additionally introduce $mRR$ which normalizes this score by the number of papers written by the reviewer. $BordaFuse$ utilises Borda-fuse as score. The three voting techniques $exp^{SUM}$, $exp^{ANZ}$ and $exp^{MNZ}$ are defined as their non-exponential forms but instead of using similarities, they apply the exponential function on similarities. 
	
	For a fixed voting technique, this step generates a ranked list $RL$ of reviewers, i.e. experts, fitting the manuscript in question. 
	
	\subsection{Step 2: Reviewer Set Assignment}
	
	Step 2 handles the right part of Figure~\ref{fig:steps}, i.e. the actual formation of reviewer sets for manuscript $M$ based on the ranked list $RL$ of reviewers generated in Step 1. We denote the top $k$ reviewers from $RL$ by $RL_{top}$; if $k = |RL|$, the first step becomes irrelevant. A smaller $k$ restricts the observed candidates in the second step drastically and is especially useful to improve runtime.\footnote{The influence of cutoff $k$ on the overall performance is evaluated in Section~\ref{eval:ae}.}
	
	We now represent documents by term-based vectors weighted with tf-idf and by topic-based vectors computed with LDA~\cite{lda}; this allows us to capture concrete terms as well as general topics of publications of reviewer candidates and the submission.\footnote{Both for tf-idf as well as LDA vector representations of documents, values in all dimensions are non-negative.} 
	Additionally, these document vector representations allow us to easily weight and combine vectors of publications without destroying their expressiveness as each vector dimension represents a single token or topic which can be present in a document to a certain extent. This starkly contrasts BERT or Doc2Vec embeddings, where single dimensions do not have a comprehensible semantics but instead the combination of all dimensions represents a document entirely.
	These tf-idf and LDA vectors can be constructed either on all parts of manuscripts or only on the technical sections, which consist of the methodology as well as the evaluation.
	
	For each reviewer $R$ this step considers the set $r_t(R,M)$ of her publications whose similarity to manuscript $M$ is not lower than a threshold $t$; i.e. $r_t(R, M) = \{P | P \in P(R) \land sim(P, M) \ge t\}$, with $t \in [0, 1]$. The threshold is utilised to define the selectivity of the research area relevant for the submission. If $t=0$ all papers of a reviewer are included, a value closer to $1$ restricts the number of papers taken into account in the second step. We assume similarities lie in $[0, 1]$.
	
	Let $rep(P, V)$ be the representation of publication $P$ as a vector of type $V \in \{L, T\}$ with $L$ representing LDA vectors and $T$ representing tf-idf vectors. Both document vector representations (DVs) can be used to compute $r_t(R,M)$, e.g. using the cosine of the corresponding vectors as similarity function.
	
	Lastly, let $P_{V, R, M, t} = \frac{\sum_{P \in r_t(R, M)} rep(P, V)}{||\sum_{P \in r_t(R, M)} rep(P, V)||_2}$ be the length normalized aggregation vector of type $V$ that combines all information on relevant publications of a reviewer $R$ with respect to $M$.
	
	We now consider all possible candidate reviewer sets of a predefined size (for example 3) and assess, for each candidate set $R_c$, its suitability with respect to the aspects defined in Section~\ref{aspects}. We prohibit reviewers in a set $R_c$ to be co-authors of each other (Aspect 2); sets that include such reviewers are not considered further, they are assigned a final $score$ of $0$. 
	In addition, we observe five different quantifiable aspects for suitability for each such set $R_c$ of reviewer candidates. These reviewers are taken from $RL_{top}$ produced in Step 1. Scores for all aspects are normalised to $[0, 1]$ with $1$ being the best and $0$ being the worst possible value.
	
	\subsubsection{Expertise E} 
	\label{sec:expertise}
	Expertise describes the relevance of the reviewers in a set to the manuscript (Aspect 3). Reviewers should have solid knowledge with terms and topics of the manuscript substantiated by numerous publications. Particularly, the submission should be similar to publications written by the reviewers~\cite{robustModel} and their number of such papers should be high. 
	Contrasting Liu et al.'s work ~\cite{robustModel} we use the number of co-authorships of reviewer candidates not as an indicator of authority but rather as an indicator of expertise.
	These conditions are measured by the following scores: 
	\[
	E_1(R_c, M, t) = \frac{\sum_{R_i \in R_c} sim(P_{L, R_i, M, t}, M)}{|R_c|}
	\]
	\[
	E_2(R_c, M, t) = \frac{\sum_{R_i \in R_c} sim(P_{T, R_i, M, t}, M)}{|R_c|}
	\]
	\[
	E_3(R_c, M, t) = \frac{\sum_{R_i \in R_c} |r_t(R_i, M)|}{|R_c| \cdot max_{R\in RL_{top}}|r_t(R, M)|}\\
	\]
	These scores are then linearly combined to the final expertise score, with $\epsilon_i\in[0, 1]$ weighting parameters and $\epsilon_1 + \epsilon_2 + \epsilon_3 = 1$:
	\[
	E(R_c, M, t) = \epsilon_1 E_1(R_c, M, t) + \epsilon_2 E_2(R_c, M, t) + \epsilon_3 E_3(R_c, M, t)
	\]
	
	\subsubsection{Authority A} 
	Reviewers should hold authority in the area the manuscript belongs to (Aspect 4). We propose two scores to measure authority: the average $h$ index of reviewers~\cite{robustModel} $h(R, M, t)$ calculated on papers relevant to the manuscript $r_t(R, M)$ (measured by $A_1$), and the average number of their obtained citations on these papers (measured by $A_2$):
	\[
	A_1(R_c, M, t) = \frac{\sum_{R_i \in R_c} h(R_i, M, t)}{|R_c| \cdot max_{R\in RL_{top}}h(R, M, t)}
	\]
	
	\[
	A_2(R_c, M, t) = \frac{\sum_{R_i \in R_c}\sum_{P_j \in r_t(R_i, M)} c(P_j)}{|R_c| \cdot \max_{R\in RL_{top}}\sum_{P\in r_t(R,M)}c(P)}
	\]
	with $c(P)$ being the number of citations a paper $P$ has obtained. These scores are then linearly combined to the final authority score, with $\alpha\in[0, 1]$ a weighting parameter:
	\[
	A(R_c,M,t) = \alpha A_1(R_c, M, t) + (1-\alpha) A_2(R_c, M, t)
	\]
	
	\subsubsection{Diversity D} 
	We define diversity as a measure to ensure that the expertise of reviewers is distributed to areas as disjunct as possible (Aspect 5). This allows for reviews to cover multiple aspects of the manuscript. The corresponding score rewards if topics in which reviewers are proficient overlap as little as possible~\cite{robustModel}: 
	\[
	D(R_c, M, t) = 1 - \frac{\sum_{R_i, R_j \in R_c, i<j} sim(P_{L, R_i, M, t}, P_{L, R_j, M})}{|R_c|\cdot (|R_c|-1)/2}
	\]
	
	\subsubsection{Interest I} 
	As research objectives of scientists change over time, interest measures the fit of reviewers and the manuscript with respect to their temporal development (Aspect 6). Interest of reviewers denotes their willingness to review submissions from certain areas~\cite{trendResInt2}. These interests change over time. If a reviewer was involved in a topic several years ago but then changed her focus, she probably no longer follows the rapid developments in the former research area. Thus she might not be willing or even able to review current submissions from this area. 
	To represent the time-aware profiles of reviewers, we combine the publications of reviewers with regards to their age to a length-normalized vector where recent papers are weighted stronger than older ones. This measure works on topical representations of documents:
	\[
	I(R_c, M, t) = |R_c|^{-1} \cdot \sum_{R_i \in R_c} sim\left(\frac{\sum_{P_j \in r_t(R_i, M)} \frac{rep(P_j, L)}{a(P_j)}}{||\sum_{P_j \in r_t(R_i, M)} \frac{rep(P_j, L)}{a(P_j)}||_2}, M\right)
	\]
	with $a(P)$ describing the age of a publication $P$ in years.
	
	\subsubsection{Seniority S} 
	\label{sec:seniority}
	In terms of seniority, reviewer sets are desirable which do not solely consist of senior researchers (Aspect 7). In the recommended group of candidates, at least one senior researcher  should be contained who is familiar with the methodology of the paper \cite{sigmod} (measured by $S_2$). Further it is desirable to have a diverse group in terms of seniority, the set should include at least one junior researcher (measured by $S_1$).
	These requisitions are modelled in the following equations:
	%
	%
	%
	\[
	S_1(R_c, M, t) = 1 - \frac{min_{ R_i \in R_c}range(R_i, M, t)}{max_{R \in RL_{top}}range(R, M, t)}
	\]
	
	\[
	S_2(R_c, M, t) =  min\left(\frac{max_{R_i \in R_c}range(R_i, M, t)}{quantile_{.75, R \in RL_{top}}{range(R, M, t)}}, 1\right)
	\]
	with $range(R, M, t) = 1 + max_{P \in r_t(R, M)}{a(P)} - min_{ P \in r_t(R, M)}$ ${a(P)}$ denoting the temporal range in which reviewer $R$ has published on topics relevant to $M$. These scores are then linearly combined to the final seniority score, with $\sigma\in[0, 1]$ a weighting parameter:
	\[
	S(R_c,M,t) = \sigma S_1(R_c, M, t) + (1-\sigma) S_2(R_c, M, t)
	\]
	\subsubsection{Final Equation}
	
	
	We combine all of these five quantifiable aspects to obtain a single score $SC$ for each reviewer sets. Good reviewer sets will have  high values in all aspects; we thus multiply the per-aspect scores:
	\begin{equation}
	\begin{split}
	SC(R_c, M, t) =& A(R_c, M, t) \cdot S(R_c, M, t) \cdot I(R_c, M, t) \\
	& \cdot D(R_c, M, t) \cdot E(R_c, M, t) 
	\end{split}
	\label{eq1}
	\end{equation}
	%
	The candidate reviewer set $R_c$ achieving the highest $SC$ is the most suitable one and recommended for the manuscript as result of Step 2. We will denote this result as $R_0$ in the experimental evaluation. 

	\section{Data sets}
	\label{ds}
	
	To evaluate our proposed reviewer set recommendation approach, we develop three novel evaluation data sets. We consider man\-u\-scripts from three different workshops and conferences of different size and thematic focus that took place in 2017, namely MOL, BTW, and ECIR. As it is practically impossible to obtain all papers submitted to a conference, we use all accepted papers as an approximation instead. Note that this might lead to non-representative topic distributions of manuscripts and unrealistically low number of manuscripts to be reviewed. Additional fuzziness is introduced since we do not distinguish between long, short and demo papers as program committees are oftentimes published in a merged form. 
	
	\subsection{Data Acquisition}
	
	We built three different data sets\footnote{Available under 
		\href{https://doi.org/10.5281/zenodo.4071874}{https://doi.org/10.5281/zenodo.4071874}. Data acquired from the manual evaluations in Section~\ref{eval} as well as templates showcasing the structure of the files are also included in the data sets.} based on data from dblp~\cite{ley}\footnote{As of January 1, 2020; \url{https://dblp.org/xml/release/dblp-2020-01-01.xml.gz}} which was merged with abstracts, citations and references from the AMiner part of the Open Academic Graph~\cite{aminer,DBLP:conf/www/SinhaSSMEHW15}\footnote{V1 from mid 2017; \url{https://www.aminer.org/open-academic-graph}} where available as well as full texts of accepted manuscripts. Information from AMiner was joined with dblp data (based on matching DOIs where available, or on matching paper titles, author names and publication years otherwise); this allowed to focus on publications from computer science or adjacent domains and to build rather precise reviewer profiles due to dblp's author disambiguation efforts, compared to using reviewer names only. Full texts of accepted manuscripts are not included in the AMiner data set but stem from pdfs collected by hand which were converted to text files using Science Parse\footnote{V2.03; \href{https://github.com/allenai/science-parse}{https://github.com/allenai/science-parse}}. 
	
	Information on program committees was either taken from conference web sites or conference proceedings. Reviewer names were manually mapped to dblp authors.\footnote{Note that the dblp data set is being revised continuously, reviewers' profiles might be imperfect due to disambiguation problems.}
	For each reviewer, we set up a list of her publications identified by their dblp keys. Here, only papers up to 2016 were taken into consideration, corresponding to a reviewer selection process in early 2017. For each of the papers the data set contains its publication year, the paper length, the CORE rank\footnote{\href{http://www.core.edu.au/}{http://www.core.edu.au/}} of the venue it was published in, the number of citations it accumulated and the average $h$ index of its authors. The concatenated title and abstract (where available) of papers needed to consist of at least three terms to be considered for the data set. Citing papers which are not contained in dblp were omitted. Thus, the number of incoming links might not necessarily represent the number of citations which publications received in the real world. This influences the number of citations and the average $h$ index.
	
	For each manuscript of our three test conferences the data sets contain a pool of possible reviewers. It consists of all members of the program committee, but excludes those with obvious conflicts of interest accessible by (former) co-authorships of authors of the manuscripts and reviewers. 
	For each of the papers published by possible reviewers, our data sets also contain tf-idf, Doc2Vec~\cite{d2v}, LDA~\cite{lda} and BERT~\cite{BERT} vector representations of its title and abstract where available. For submitted manuscripts, these four kinds of document representation are contained for the full text as well as only the research sections of the paper (which consist of all sections excluding the abstract, introduction, related work, conclusion, references and acknowledgements). The textual content of the papers is not contained. We consider only English documents for the construction of our data sets.
	
	\subsection{Document Representations}
	
	We calculated the document frequencies of words for tf-idf on unstemmed titles of all publications contained in dblp up to 2016 concatenated with abstracts from  AMiner where available which were written in English. In total we used 2,940,996 documents.
	The final tf-idf vectors are calculated for unstemmed textual data available in the respective data sets including all papers of reviewers and submitted manuscripts.
	
	For the construction of BERT~\cite{BERT} vectors, we used the base pretrained uncased model.\footnote{We utilised the BERT implementation and model provided by \href{https://huggingface.co/transformers/model_doc/bert.html}{https://huggingface.co/transformers/model\_doc/bert.html}.} Since the BERT implementation used is only able to process input vectors of at most 512 tokens, documents were cut at punctuation marks or after half of the tokens if sentences were still too long. A sliding window was used to always input two consecutive sentences to maintain as much context as possible. The model consists of overall twelve hidden layers each having 768 features. The last four layers from these twelve layers were concatenated for each token and averaged over all tokens to receive vectors of length 4 layers$\times$768 features = 3072 dimensions for each publication.~\cite{kreutz}
	
	Weights for Doc2Vec~\cite{d2v} are trained on the English Wikipedia corpus from 1st February 2020\footnote{\href{https://dumps.wikimedia.org/enwiki/20200201/}{https://dumps.wikimedia.org/enwiki/20200201/}}. We refrained from using Doc2Vec on a stemmed corpus as this preprocessing is no prerequisite for achieving good results~\cite{d2v}. 
	We trained two Doc2Vec models, one distributed bag of words (DBOW) and one distributed memory (DM) model, so that resulting vectors consist of 300 dimensions each. This size was proposed by Lau and Baldwin~\cite{numdim} for general-purpose applications.\footnote{We utilised the Doc2Vec implementation provided by \href{https://radimrehurek.com/gensim/models/doc2vec.html}{https://radimrehurek.com/gensim/models/doc2vec.html}.}~\cite{kreutz}
	
	For LDA~\cite{lda} we again used the 2,940,996 documents which we already utilised for the computation of the document frequency in tf-idf. This procedure ensured the computed topics were from the area of computer science. The number of topics was set to 100 resulting in the same number of dimensions for vector representations of manuscripts and publications.\footnote{We utilised the LDA implementation provided by \href{https://radimrehurek.com/gensim/models/ldamodel.html}{https://radimrehurek.com/gensim/models/ldamodel.html}.}~\cite{kreutz}
	
	\subsection{MOL'17, BTW'17 and ECIR'17}
	
	\subsubsection*{MOL'17}
	The data set contains 12 manuscripts in English language which were accepted at \textit{Meeting on the Mathematics of Language '17}, 22 program committee members and their papers in dblp. We excluded extended abstracts. No distinction between different paper types and program committees was made. On average each manuscript has 21 possible reviewers which do not have conflicts of interests. This data set represents a small biannual international conference with a different focus than the other two data sets.

	\subsubsection*{BTW'17}
	The data set contains 36 manuscripts in English language which were accepted at \textit{Datenbanksysteme für Business, Technologie und Web '17} (the German database conference), 56 program committee members and their papers in dblp. We again excluded extended abstracts. No distinction between different paper types was made but the program committees members are split in scientific, industry and demo paper committee. On average each manuscript has 47.78 possible reviewers which do not have conflicts of interests. This data set represents a medium sized biannual national conference with several lesser-known reviewers.
	
	\subsubsection*{ECIR'17}
	The data set contains 80 manuscripts in English language which were accepted at \textit{European Conference on Information Retrieval '17}, 
	151 program committee members and their papers in dblp. A distinction between full-paper meta-reviewers, full-paper program committee, short paper program committee and demonstration reviewers was made. On average each manuscript has 141.35 possible reviewers which do not have conflicts of interests. This data set represents a medium to large annual European conference attributed with CORE rank A and mostly well-known reviewers.
	
	\section{Evaluation}
	\label{eval}
	
	In our experiments, we solely focus on sets consisting of three reviewers even though our approach is applicable for different numbers of reviewers per manuscript as well. This number was chosen as a widespread norm~\cite{toronto} to reduce the dimensionality of further evaluation steps.
	We evaluate our approach on the three introduced data sets MOL'17, BTW'17 and ECIR'17 where we disregard the different manuscript and committee types. By observing the performance of our approach in venues of different sizes, we strive to make assumptions on its general applicability. We use Cosine similarity as similarity measure. This ensures similarity values in $[0, 1]$ for Step 2 as tf-idf and LDA document vector representations hold non-negative values for all dimensions.
	For the voting techniques 
	of the algorithm we run tests with $n \in \{5, 10\}$ and $\delta \in \{0, .25, .5, .9\}$. 
	
	For all significance tests, we use a $p$-value of .05. We evaluate the normal distribution of values using Kolmogorov-Smirnov tests and test the homogeneity of variances with Levene's tests.
	All depicted values are rounded on four decimal places.
	
	\subsection{Hypotheses}
	
	Considering the overall challenges and goals of RevASIDE, we investigate the following six hypotheses: 
	
	\begin{itemize}
		\item [$H_1$] Step 1 is useful for the expert search task.
		\item [$H_2$] Usage of more advanced document vector representations leads to significantly better overall results for Step 1 compared to more basic ones.
		\item [$H_3$] Utilisation of different document vector representations, voting techniques, cutoff values $k$ of the result list $RL$, content types and thresholds $t$ leads to significantly different overall RevASIDE scores and values for the five quantifiable aspects in Step 2.
		\item [$H_4$] Utilisation of the full texts of manuscripts leads to worse overall results than restriction of the manuscripts' content to the technical sections in Step 2.
		\item [$H_5$] The conduction of Step 1 is profitable for Step 2.
		\item [$H_6$] Results of Step 2 are confirmed by human assessment, thus RevASIDE is useful for the reviewer set assignment task.
	\end{itemize}

	\subsection{Step 1: Expert Search Task}
	
	In this part of the evaluation we intend to assess hypotheses $H_1$ of Step 1 being useful for the expert search task and $H_2$ of utilisation of more advanced DVs producing better results.
	
	\begin{table*}[t]
		\caption{Mean average precision@10 (MAP), precision@10 (P@10) and nDCG@10 (nDCG) for all combinations of voting techniques (VT) and document vector representations of manuscripts from BTW'17 (upper half) and ECIR'17 (lower half). Best combination in BTW'17: tf-idf + $SUM$ (short $b_1$). Best combinations in ECIR'17: tf-idf + $MNZ$ (short $e_1$), DBOW + $Votes_{\delta=.5}$ (short $e_2$).
			Column sig diff gives information on whether or not MAP (m), P@10 and nDCG (n) significantly differ between the different DVs. If $\checkmark$, all three measures are significantly different.}
		\label{tab:e1btw}
		\centering
		\begin{tabular}{l"l|l|l"l|l|l"l|l|l"l|l|l"l}
			data set  &  \multicolumn{13}{c}{\textbf{BTW'17}}\\
			
			DV & \multicolumn{3}{c"}{tf-idf}& \multicolumn{3}{c"}{DM} & \multicolumn{3}{c"}{DBOW} & \multicolumn{3}{c"}{BERT} & sig diff\\ \hline
			VT$\backslash$measure& MAP & P@10 & nDCG & MAP & P@10 & nDCG & MAP & P@10 & nDCG & MAP & P@10&  nDCG \\ 
			\hline 
			
			\hline
			$Votes_{\delta=0}$ & .1705	&	.34&	.3677	&	.1732	&	.345&	.3706	&	.1705	&	.34&	.3677	&	.1705	&	.34&	.3677 & \\
			$Votes_{\delta=.25}$ & .0712	&	.2&	.1881	&	.2056	&	.36&	.4017	&	.1714	&	.335&	.3685	&	.1705	&	.34&	.3677 & $\checkmark$\\
			$Votes_{\delta=.5}$ & .0712	&	.2&	.1881	&	.1584	&	.25&	.3088	&	.1966	&	.325&	.373&	.1705	&	.34&	.3677 & $\checkmark$\\
			$Votes_{\delta=.9}$ & .0712	&	.2&	.1881	&	.0712	&	.2&	.1881	&	.0712	&	.2&	.1881	&	.1766	&	.35&	.3701 & $\checkmark$\\
			$SUM$ & \textbf{.2947}	&	\textbf{.42}&	\textbf{.4923}	&	.1816	&	.345&	.3721	&	.1749	&	.34&	.3722	&	.168&	.34&	.3635 & \\
			$AVG$ & .2612	&	.385&	.4246	&	.1222	&	.265&	.2754	&	.1682	&	.345&	.345&	.0604	&	.16&	.1593 & $\checkmark$\\
			$MNZ$ & .273&	.41&	.4761	&	.1787	&	.345&	.3777	&	.1755	&	.345&	.3725	&	.1704	&	.34&	.367 & \\
			$SUM_{n=5}$ & .0303	&	.1&	.1007	&	.043&	.15&	.1398	&	.0697	&	.18&	.1827	&	.0385	&	.115&	.1113 & m, n\\
			$SUM_{n=10}$ & .0329	&	.08&	.0891	&	.0421	&	.135&	.1321	&	.0659	&	.175&	.1813	&	.0231	&	.095&	.0872 & $\checkmark$\\
			$MIN$ & .0364	&	.155&	.1194	&	.0301	&	.135&	.1023	&	.0391	&	.145&	.1316	&	.0168	&	.08&	.0622 & \\
			$MAX$ & .2779	&	.39&	.4589	&	\textbf{.2872}	&	\textbf{.405}&	\textbf{.4812}	&   \textbf{.256}&	\textbf{.395}&	\textbf{.4517}	&	.2162	&	.34&	.3758 & \\
			$RR$ & .0949	&	.235&	.242&	.1027	&	.265&	.2632	&	.1005	&	.25&	.2386	&	\textbf{.2326}	&	\textbf{.365}&	\textbf{.4311} & $\checkmark$\\
			$mRR$ & .0519	&	.165&	.1505	&	.0613	&	.2&	.1882	&	.0683	&	.185&	.1857	&	.1757	&	.33&	.3614 & $\checkmark$\\
			$BordaFuse$ & .1545	&	.325&	.3405	&	.1385	&	.305&	.3192	&	.12&	.275&	.2768	&	.1633	&	.345&	.3459 & \\
			$exp^{SUM}$ & .1705	&	.34&	.3677	&	.1764	&	.345&	.3756	&	.1725	&	.34&	.3695	&	.168&	.34&	.3635 & \\
			$exp^{AVG}$ & .2612	&	.385&	.4246	&	.1248	&	.265&	.2759	&	.171&	.34&	.3458	&	.0589	&	.16&	.1542 & $\checkmark$\\
			$exp^{MNZ}$ & .1705	&	.34&	.3677	&	.1761	&	.345&	.3752	&	.171&	.34&	.3681	&	.1708	&	.34&	.3679 & \\
			
			\hline 
			
			\hline
			data set  &  \multicolumn{13}{c}{\textbf{ECIR'17}}\\
			
			DV & \multicolumn{3}{c"}{tf-idf}& \multicolumn{3}{c"}{DM} & \multicolumn{3}{c"}{DBOW} & \multicolumn{3}{c"}{BERT} & sig diff\\ \hline
			VT$\backslash$measure& MAP & P@10 & nDCG& MAP & P@10 & nDCG & MAP & P@10 & nDCG & MAP & P@10&  nDCG \\ 
			\hline 
			
			\hline
			$Votes_{\delta=0}$ & .1116	&	.45	&	.4748	&	.1132	&	.455	&	.4784	&	.1116	&	.45	&	.4748	&	.1116	&	.45	&	.4748	&\\ 
			$Votes_{\delta=.25}$ & .031	&	.21	&	.2095	&	.1308	&	.485	&	.5245	&	.1195	&	.48	&	.4937	&	.1116	&	.45	&	.4748	& $\checkmark$\\
			$Votes_{\delta=.5}$ & .0317	&	.21	&	.2147	&	.1239	&	.43	&	.4733	&	\textbf{.164}	&	\textbf{.555}	&	\textbf{.5992}	&	.1116	&	.45	&	.4748	& $\checkmark$\\
			$Votes_{\delta=.9}$ & .0317	&	.21	&	.2147	&	.0317	&	.21	&	.2147	&	.0317	&	.21	&	.2147	&	\textbf{.1252}	&	\textbf{.48}	&	\textbf{.5081}	& $\checkmark$\\
			$SUM$ & .1217	&	.475	&	.4789	&	.1283	&	.49	&	.5173	&	.124	&	.475	&	.5007	&	.115	&	.46	&	.482	&\\
			$AVG$ & .0664	&	.315	&	.3129	&	.047	&	.285	&	.2639	&	.0567	&	.33	&	.3167	&	.0349	&	.19	&	.1843	& $\checkmark$\\
			$MNZ$ & \textbf{.1647}	&	\textbf{.545}	&	\textbf{.5908}	&	.124	&	.475	&	.5012	&	.1163	&	.46	&	.4826	&	.1132	&	.455	&	.4788	&\\
			$SUM_{n=5}$ & .0309	&	.19	&	.1792	&	.0202	&	.135	&	.1353	&	.0286	&	.185	&	.1679	&	.0431	&	.175	&	.1736	&\\
			$SUM_{n=10}$ & .0284	&	.16	&	.1691	&	.0243	&	.12	&	.1311	&	.0297	&	.15	&	.1615	&	.0473	&	.165	&	.1731	&\\
			$MIN$ & .031	&	.23	&	.2004	&	.0181	&	.165	&	.149	&	.0206	&	.14	&	.1392	&	.0274	&	.17	&	.1588	&\\
			$MAX$ & .1205	&	.41	&	.4429	&	\textbf{.1496}	&	\textbf{.535}	&	\textbf{.5449}	&	.1525	&	.535	&	.565	&	.0717	&	.38	&	.362	& $\checkmark$\\
			$RR$ & .0535	&	.275	&	.2858	&	.084	&	.375	&	.3926	&	.0651	&	.37	&	.3451	&	.0765	&	.39	&	.3693	&\\
			$mRR$ & .0141	&	.135	&	.1186	&	.0311	&	.23	&	.2197	&	.0318	&	.24	&	.2299	&	.0264	&	.185	&	.1913	& $\checkmark$\\
			$BordaFuse$ & .099	&	.445	&	.4393	&	.0957	&	.425	&	.427	&	.0921	&	.415	&	.4137	&	.1014	&	.43	&	.4404	&\\
			$exp^{SUM}$ & .1116	&	.45	&	.4748	&	.117	&	.465	&	.485	&	.116	&	.46	&	.4828	&	.115	&	.46	&	.4818	&\\
			$exp^{AVG}$ & .0658	&	.315	&	.3164	&	.049	&	.295	&	.2725	&	.0574	&	.33	&	.3171	&	.038	&	.19	&	.1871	& $\checkmark$\\
			$exp^{MNZ}$ & .1116	&	.45	&	.4748	&	.115	&	.46	&	.482	&	.115	&	.46	&	.482	&	.1132	&	.455	&	.4788	&\\ 
		\end{tabular}
	\end{table*}
	
	We randomly selected 20 manuscripts from each of the BTW'17 and ECIR'17 data sets. The manuscripts are represented by their full texts, the profiles of reviewers are represented by their papers' titles and abstracts where available. To create a ground-truth of relevant reviewers, the top 10 reviewer candidates are computed with all 13 (17 with variants) voting techniques and combined. The resulting pools of reviewers for each manuscript from the BTW'17 data set contained 48.35 entries on average and 101.5 entries on average for manuscripts from ECIR'17. In the former case, about all possible reviewers were contained in the respective lists contrasting the ECIR'17 lists which contain a lower percentage of possible reviewers.
	Unfortunately, a more extensive manual evaluation with more manuscripts would not be feasible.  
	
	The manuscripts' title and abstract as well as the potential reviewers and a link to their dblp profile were presented to an independent senior researcher in the field who evaluated the reviewers in terms of appropriateness for the given manuscript. For the manual evaluation of relevance, only papers up to 2016 of reviewers were considered. The expert was not aware which method retrieved which reviewers. If the expert observed missing relevant reviewers, they were also included in the ground-truth. In BTW'17, each paper has 10.05 relevant reviewers on average (min=5, max=14, median=10, standard deviation=2.762). In ECIR'17, each paper has 27.2 relevant reviewers on average (min=3, max=55, median=25, standard deviation=13.5671).
	On average, a reviewer from the program committee is relevant for 3.5893 manuscripts for BTW'17 and 3.1813 manuscripts for ECIR'17.
	
	We report result quality with three established metrics, examining the first 10 retrieved reviewers of each method. Precision@10 measures the fraction of the top-10 recommended reviewers that were actually relevant. Non-interpolated mean average precision@10 (MAP) averages the precision at ranks where a relevant reviewer appears, using a precision of 0 for each relevant reviewer not appearing in the result list. Normalized cumulative discounted gain (nDCG)~\cite{DBLP:journals/tois/JarvelinK02} aggregates relevance of all reviewers appearing in the result, but with a logarithmic discount for later ranks; this follows the intuition that later ranks are less important to a user than earlier ranks. In addition, it normalizes this aggregation by the cumulative discounted gain achieved by an ideal ranking where all relevant reviewers appear in front, thus showing how close the result is to an optimal result and allowing to compare across different queries with different numbers of relevant results.
	
	The upper part of Table~\ref{tab:e1btw} shows result quality for all combinations of document vector representation and voting technique for the twenty manuscripts from BTW'17. $Votes_{\delta=0}$ is exactly the same for each document vector representation as this voting technique solely considers the number of papers of reviewer candidates and not their similarity with query manuscripts. The lower part of Table~\ref{tab:e1btw} shows the same for the twenty manuscripts from ECIR'17.
	
	In BTW'17, each paper has 2.7801 relevant reviewers per combination of VT and DV on average, in ECIR'17 this value is significantly (Mann-Whitney $U$ test) higher (3.4838). 
	These assessments lead to the assumption of the VTs and DVs presented here being useful for the expert search task and therefore verifying $H_1$.
	
	We found significant (Kruskal-Wallis $H$ tests) differences between the four DVs for several voting techniques, but not for all of them (see rightmost column of Table~\ref{tab:e1btw}). The more advanced document vector representations Doc2Vec and especially BERT did not achieve better results than tf-idf.
	
	The best voting techniques seem to depend on the data set and the utilised document vector representation. BERT performs worse than both tf-idf and the Doc2Vec models. 
	Usage of tf-idf and DM achieves comparable results for the best performing VTs for BTW'17; for ECIR'17, tf-idf and DBOW with their respective best VTs result in similar values. 
	BERT seems to generalise the concepts of papers too much such that the VTs cannot clearly distinguish between relevant and non-relevant reviewers. This is underlined by the fact that three versions of $Votes_{\delta}$ generate the same values for MAP, P@10 as well as nDCG. Tf-idf has high selectivity and is able to identify experts versed in the exact same techniques described in a manuscript.
	Hence, hypothesis $H_2$ of more sophisticated VRs being more suitable than basic VRs is rejected. 
	
	For the ECIR'17 data set, P@10 and nDCG are higher than for BTW'17. This might be caused by ECIR'17 having higher overall numbers of reviewers as well as more relevant reviewers per manuscript. This disadvantages the smaller BTW'17 data set. 

	\subsection{Step 2: Reviewer Set Assignment Task}
	The evaluation of Step 2 of our algorithm consists of a quantitative and a qualitative evaluation. In Equation~\ref{eq1} we set $\epsilon_{1} = \epsilon_{2} = \epsilon_{3} = \frac{1}{3}$ and $\alpha = \sigma = .5$.
	
	As a first baseline $B_{t3}$, the three highest ranked reviewers in the ranked list $RL$ for each VT and DV are considered as a reviewer set for a manuscript. Such an approach is common in reviewer set recommendation~\cite{robustModel,mlClass}. Our second baseline $B_{tr}$ chooses three random reviewers from $RL_{top}$. Our third baseline $B_{r}$ chooses three random reviewers from the whole program committee, excluding only those with a conflict of interest. For the latter, we cap values of $E_3$, $A_1$ and $A_2$ at 1.\footnote{Reviewers from $B_{r}$ are possibly not contained in $RL_{top}$ and thus could theoretically produce values $>1$ for the three partial aspects. The score for this baseline is still calculated based on maxima of papers of relevant candidate reviewers.}
	
	We experiment with cutoffs $k$ of reviewers in $RL$ to generate $RL_{top}$ at position 10 and 20 after Step 1 and without cutoff, i.e. all reviewers without conflicts of interests for the manuscripts were utilised as a comparison to evaluate the usefulness of Step 1. 
	If we do not restrict the number of candidate reviewers, i.e. $|RL| = k$, the voting technique used in Step 1 (which determines the reviewer candidates considered in Step 2) becomes irrelevant for Step 2 but still influences the creation of the 
	baselines.
	We also experiment with different thresholds $t \in \{0, .25, .5, .9\}$.
	
	We divide the manuscripts in non-technical and research sections to better estimate their true content. Non-technical sections include abstract, introduction, related work, conclusion, acknowledgements and references. Research sections are all other parts. 
	We compare the effect of using the full text in Step 2 to using only the content of research sections. Profiles of reviewers are represented by their papers' titles and abstracts where available which are similar enough (threshold $t$) to the query manuscript.
	
	\subsubsection{Quantitative Evaluation}
	\label{eval:ae}
	
	In this part of the evaluation we focus on understanding the influence of the different factors of our approach and prepare the qualitative evaluation by identifying the combinations achieving the highest scores.
	In this context we intend to assess 
	hypotheses $H_3$ and $H_4$ as well as $H_5$ which observes the usefulness of Step 1.
	
	In these experiments, for each combination of document vector representation in Step 1, voting technique, cutoff of relevant reviewers utilised in Step 2, similarity threshold $t$ in Step 2 as well as used content type (CT) in Step 2 we observe the following result types (RT): the three baselines ($B_{t3}$, $B_{tr}$, $B_{r}$) and the best result returned by RevASIDE ($R_0$).
	
	\begin{table}[t]
		\caption{Significant differences between the groups in $SC$ as well as the five quantifiable aspects by data sets MOL'17 (m), BTW'17 (b) and ECIR'17 (e).}
		\label{tab:e2groupBy}
		\centering
		\begin{tabular}{l|l|l|l|l|l|l}
			grouped by & $SC$ & $A$ & $S$ & $I$& $D$ & $E$ \\ \hline
			DV                  & mb             & mb           & b   & e            & mbe & mbe \\
			VT                  & mb             & mbe          & mbe & mbe          & mbe & mbe \\
			$k$        & mbe            & mbe          & mbe & mbe          & mbe & mbe \\
			CT                  & mbe            &              &     & mbe          &     & mbe \\
			$t$ in Step 2       & mbe            & mbe          & mbe & mbe          & mbe & mbe \\
			RT                  & mbe            & mbe          & mbe & mbe          & mbe & mbe
		\end{tabular}
	\end{table}{}
	
	\begin{table*}[t]
		\caption{Configuration (conf), DV, VT, $RL_{top}$ cutoff value $k$, utilised content type and threshold $t$ resulting in the highest average scores and corresponding values for $A$, $S$, $I$, $D$ as well as $E$ per data set and result type.}
		\label{tab:best_comb}
		\centering
		\begin{tabular}{l|l|l|l|l|l|l|l"l|l|l|l|l|l}
			conf &data set & RT & DV & VT & CT & $k$ & $t$ & $SC$ & $A$&$S$& $I$ & $D$ & $E$\\ 
			\hline 
			
			\hline
			$c_1$            & MOL'17                        & $R_0$                   & BERT                    & $MIN$                   & full                    & 20                                & .5                       & \textbf{.053}  & .6675 & .8696 & .5277 & .3783 & .4614 \\
			$c_2$            & MOL'17                        & $B_{t3}$             & BERT                    & $Votes_{\delta=0}$           & full                    & 20                                & .25                      & .0439                           & .5972 & .8696 & .5283 & .3109 & .5056 \\
			$c_3$            & MOL'17                        & $B_{tr}$              & DBOW                    & $exp^{AVG}$             & full                    & 10                                & .25                      & .0348                           & .69   & .8158 & .5158 & .3014 & .3738 \\ 
			
			$c_4$            & MOL'17                        & $B_{r}$             & DM                      & $BordaFuse$             & full                    & 10                                & .5                       & .0251                           & .4642 & .8199 & .5408 & .2654 & .4467 \\ \hline
			$c_5$            & BTW'17                        & $R_0$                   & BERT                    & $MIN$                   & full                    & 20                                & .5                       & \textbf{.0528} & .5827 & .9935 & .5243 & .555  & .3152 \\
			$c_6$            & BTW'17                        & $B_{t3}$             & BERT                    & $SUM_{n=5}$             & full                    & 10                                & .5                       & .0218                           & .4106 & .887  & .6078 & .3332 & .2891 \\
			$c_7$            & BTW'17                        & $B_{tr}$              & tf-idf                  & $SUM_{n=10}$            & full                    & 10                                & .5                       & .0303                           & .826  & .7274 & .5559 & .2298 & .3884 \\
			
			$c_8$            & BTW'17                        & $B_{r}$             & BERT                    & $MIN$                   & full                    & 10                                & .25                      & .0193                           & .3797 & .8963 & .5192 & .3743 & .2693 \\ \hline
			$c_9$            & ECIR'17                       & $R_0$                   & BERT                    & $mRR$                   & full                    & 20                                & .9                       & \textbf{.0438} & .6348 & .8077 & .6614 & .3162 & .4148 \\
			$c_{10}$           & ECIR'17                       & $B_{t3}$             & BERT                    & $mRR$                   & full                    & 10                                & .9                       & .0192                           & .5312 & .7315 & .6704 & .2038 & .3471 \\
			$c_{11}$           & ECIR'17                       & $B_{tr}$              & DM                      & $SUM_{n=5}$             & full                    & 10                                & .5                       & .0319                           & .9517 & .6077 & .6675 & .1977 & .4309\\
			
			$c_{12}$           & ECIR'17                       & $B_{r}$             & DBOW                    & $BordaFuse$             & full                    & 10                                & .5                       & .0171                           & .5228 & .7271 & .6715 & .1451 & .4581 \\
		\end{tabular}
	\end{table*}{}
	
	We test for significant differences between groups of experiments to determine which factors really influence the overall score $SC$ (as computed by Equation~\ref{eq1}) and the five quantifiable aspects introduced in Sections \ref{sec:expertise} to \ref{sec:seniority}. 
	Kruskal-Wallis $H$ tests are used for the following experiments since in most of our observed cases, data is not normally distributed in the different groups or variances are not homogeneous. Table~\ref{tab:e2groupBy} indicates between which groups of experiments we found significant differences in the scores or the five quantifiable aspects. We observe 1,632 (4 DVs$\times$17 VTs$\times$3 cutoffs $k$$\times$2 CTs$\times$4 $t$ in Step 2) experimental setups per data set. 
	Experiments were grouped by document vector type such that there were four groups of experiments, ones using tf-idf in the first step, ones using Doc2Vec DM, ones using Doc2Vec DBOW and ones using BERT document vector representations. Grouping by VT in Step 1 results in 17 different groups of experiments. 
	When experiments are grouped by the number of observed candidates $k$ three different groups result. 
	When grouping by content type, two groups of experiments result, ones which utilise the full text in Step 2 and ones utilising only the research sections of the query manuscript. 
	Grouping by the threshold value $t$ in Step 2 results in four different groups. 
	Lastly, grouping by RT produces four groups containing experiments of types $B_{t3}$, $B_{r}$, $B_{tr}$ and $R_0$. 
	
	DV does influence some aspects significantly but overall, the scores of the ECIR'17 data set are not significantly influenced by it. 
	VT significantly influences the five aspects for all data sets as well as the score for the two smaller ones. 
	The content type which is utilised in Step 2 is significantly influential for values for all data sets except for authority, diversity, and seniority. These values are not calculated by directly utilising the query manuscript and therefore are not influenced by the content type. 
	The cutoff value $k$ which is chosen for $RL_{top}$, the threshold value $t$ as well the result types significantly influence the results in all three data sets.
	From these observations we derive the overall validity of hypothesis $H_3$.
	
	Table~\ref{tab:best_comb} shows the best combinations of DV, VT, cutoff values, content type and threshold, measured in terms of the highest overall average scores for $R_0$ and the three baselines $B_{t3}$, $B_{r}$ and $B_{tr}$ for each of the three data sets.
	$SC$ is calculated with Equation~\ref{eq1} and can take take values between $0$ and $1$ with $1$ being the best. As it is multiplicative, a score of $.05$ can be reached if e.g. values of all quantifiable aspects $A$, $S$, $I$, $D$, and $E$ are around $.55$.
	
	$R_0$ achieves the highest $SC$ results for each data set. This, together with the significant differences between result types observed in the previous experiment (see Table~\ref{tab:e2groupBy}), leads to the conclusion that RevASIDE produces significantly higher average $SC$ scores than the baselines. This applies to all three different sized data sets which highlights the general applicability of our approach. 
	
	Utilising full texts of query manuscripts yields better results than only taking the research sections into account. This leads to the rejection of hypothesis $H_4$. 
	
	The restriction of $RL_{top}$ to $k = 10$ leads to the best average scores for MOL'17 and ECIR'17; for BTW'17, no restriction of $RL_{top}$ leads to the highest average scores (not depicted in the table).
	This indicates that the reduction of the number of considered reviewers for Step 2 (and therefore the entirety of Step 1) is a major factor in small and large data sets. It also decreases the overall computation time which in general verifies $H_5$. MOL'17 as well as ECIR'17 represent relatively focused areas while BTW'17 is more diverse. For focused data sets it suffices to regard the few most relevant reviewers to compose a suitable set but for a diverse conference, it seems more reviewers need to be considered.
	When grouping all 1,632 experiments by voting technique and threshold, the highest average scores for MOL'17 are achieved by $exp^{SUM}$ and .5; for BTW'17 $SUM_{n=5}$ and .25; and for ECIR'17 $SUM_{n=10}$ and .9. BERT is the DV which on average performs best for each data set. They outperform the other VTs and thresholds on average but do not appear as a combination in Table~\ref{tab:best_comb} under the overall best configurations.
	Remarkably, the best results for $R_0$ in MOL'17 as well as BTW'17 were achieved by the same combination of DV, VT, CT, $k$ as well as $t$. The combination of BERT with $MIN$ or $mRR$ did not achieve any good results in our manual evaluation of Step 1 but did prove to be useful in Step 2.
	
	The highest scores for MOL'17 (.0516), BTW'17 (.0462) as well as ECIR'17 (.0399) for the best performing combinations from Step 1 of our approach ($b_1$: tf-idf + SUM, $e_1$: tf-idf + $MNZ$, $e_2$: DBOW + $Votes_{\delta=.5}$) are independent of DV and VT as they are achieved by $|RL_{top}| = |RL|$. The threshold $t$ is set to .5. These results cannot surpass the best configurations from Table~\ref{tab:best_comb} for the same data but also do not significantly differ from them.
	
	For BTW'17 as well as ECIR'17, we found no significant correlation between the scores produced by the twelve (eleven as $c_1$ = $c_5$) best configurations from Table~\ref{tab:best_comb} for $R_0$ and the number of relevant reviewers per manuscript for the forty manuscripts observed in the evaluation of Step 1 with Kendall's $\tau_B$.
	
	We want to point to the fact that some of the DVs and VTs present in Table~\ref{tab:best_comb} achieve low results in the evaluation of Step 1 (for BTW'17 .0168 to .171 in MAP, .08 to .34 in P@10 and .0622 to .3677 in nDCG; for ECIR'17 .0264 to .1116 in MAP, .135 to .45 in P@10 and .1353 to .4748 in nDCG). This hints at possible problems with aspects with opposing objectives which will be regarded in depth in the following qualitative evaluation.
	
	\subsubsection{Qualitative Evaluation}
	
	In this part of the evaluation we assess hypothesis $H_6$ which covers the manual assessment of the sets resulting from Step 2 and RevASIDE's overall usefulness.
	
	\begin{table*}[t]
		\caption{Average positions (pos) sets computed by the different configurations (conf) were ordered to in the qualitative evaluation as well as the average number of relevant reviewers (\#rel) and the average position of entries from the different RTs per set.}
		\label{tab:mane2}
		\centering
		\begin{tabular}{l"l|l"l|l"l|l"l|l"l|l"l|l"l|l"l|l}
			data set  &  \multicolumn{8}{c"}{\textbf{BTW'17}} &  \multicolumn{8}{c}{\textbf{ECIR'17}}\\ 
			result type & \multicolumn{2}{c"}{$R_0$} & \multicolumn{2}{c"}{$B_{t3}$} & \multicolumn{2}{c"}{$B_{tr}$} & \multicolumn{2}{c"}{$B_{r}$} & \multicolumn{2}{c"}{$R_0$} & \multicolumn{2}{c"}{$B_{t3}$} & \multicolumn{2}{c"}{$B_{tr}$} & \multicolumn{2}{c}{$B_{r}$}\\ \hline
			avg. position $\forall$ conf& \multicolumn{2}{l"}{\textbf{
					2.3292}} & \multicolumn{2}{l"}{2.9083}& \multicolumn{2}{l"}{2.4083} &\multicolumn{2}{l"}{2.3373}& \multicolumn{2}{l"}{\textbf{2.1454}} & \multicolumn{2}{l"}{2.7272} & \multicolumn{2}{l"}{2.4818} & \multicolumn{2}{l}{2.5818}\\\hline
			conf$\backslash$measure  & \#rel & pos &\#rel & pos & \#rel & pos & \#rel & pos  & \#rel & pos &\#rel & pos & \#rel &  pos & \#rel & pos\\ 
			\hline 
			
			\hline
			$c_1$ = $c_5$ & .3& 2.05& 0& 3.65 & .25& 2.35 & .65& 1.95& .6 & 2.4  & .2 & 3.2  & .6 & 2.3 & .4  & 2.1\\
			$c_2$ & .75& 2.8& 1.1& 2.1& .85& 2.2& .55& 2.9& 1  & 2.2 & 1.6 & 1.8 & 1.6 & 2.0 & .7  & 3.7\\
			$c_3$ & 1.1& 1.95 & .85& 3 & 1.1& 2.25 & .65& 2.8& .7  & 2.3 & .7 & 2.2 & .7 & 2.6 & .7  & 2.6\\
			$c_4$ & .6& 2.55 & .85& 2.25 & .55& 2.65& .9& 2.55& 1 & 1.6 & 1.5 & 1.9 & .8 & 2.8 & .5  & 3.7\\\hline
			$c_6$ & .4& 2.5& .25& 2.65& .5& 2.5& .55& 2.35& .2 & 2.8 & .4 & 2.6 & .4 & 2.7 & .3  & 1.9\\
			$c_7$ & .4& 2.6& .25& 2.95& .25& 2.65& .65& 1.75& .7 & 2 & .5 & 3.3 & .6 & 2.5 & .4  & 2.2\\
			$c_8$ & .2& 2.1& 0& 4 & .3& 2.25 & .5& \textbf{1.65}& .6 & 1.8 & .2 & 3.7 & .3 & 2.7 & .8  & 1.8\\\hline
			$c_9$ & .8&  2.35& 1.05& 2.3 & .85& 2.4& .6& 2.95& .5 & \textbf{1.5} & .5 & 3.2 & .5 & 2.5 & .2  & 2.8\\
			$c_{10}$ & 1.05 & 2.25& 1.05& 2.25 & .95& 2.65 & .7& 2.8& .5 & 2.4 & .5 & 2.9 & .5 & 2.7 & .7  & 2.0\\
			$c_{11}$ & .65& 2.3& .3& 3.4 & .55& 2.4 & 1& 1.9& .5 & 2.1 & .3 & 3.4 & .3 & 2.6 & .7  & 1.9\\ 
			$c_{12}$ & .65& 2.45 & .7& 2.7 & .8& 2.25 & .65& 2.5& 1 & 2.5 & 1.4 & 1.8 & 1.4 & 1.9 & .6  & 3.7\\
			
		\end{tabular}
	\end{table*}{}
	
	In our first qualitative evaluation of Step 2, we examine the eleven (as $c_1 = c_5$) configurations which performed best for the different result types from the three data sets (see Table~\ref{tab:best_comb}) in the quantitative evaluation. For the forty (twenty from ECIR'17 and twenty from BTW'17) documents which were used in the first manual evaluation, we compute lists of four reviewer sets for all  configurations, consisting of one reviewer set produced by each of the three baselines as well as $R_0$.
	We present the lists to an expert who then ranks the four entries according to suitability for the query manuscript from 1 (best) to 4 (worst), with the option of ties if two or more entries are equally suitable. 
	Table~\ref{tab:mane2} shows the average ranks of the result types in the evaluated lists for the two data sets, their average number of relevant reviewers per configuration and the average positions that entries from a specific RT achieved.
	
	For BTW'17, the combination achieving the best results is $c_{8}$ (BERT, $MIN$, $k$ = 10, $t$ = .25) and (surprisingly) $B_{r}$. 
	For ECIR'17, the combination achieving the best results came from configuration $c_9$ (BERT, $mRR$, $k$ = 20, $t$ = .9) and $R_0$. 
	
	Overall, $R_0$ achieves the best results out of all combinations and data sets.  $B_{r}$ generates the best results for BTW'17, but highly depends on the configuration as it also achieves considerably bad results, especially for the ECIR'17 data set. Although the results are greatly influenced by the configuration, $R_0$ performs consistently well in general.
	The combination of configuration and result type achieving the highest number of mean relevant reviewers per data set is not the one achieving the best results in terms of positions, e.g. ECIR'17 + $c_{c}$ + $B_{tr}$. This leads to the conclusion that it is not sufficient to consider only topical relevance in determining the most suitable combination.
	In both data sets, the RT achieving the best average positions is $R_0$.
	
	As data was not normally distributed in the different groups for both data sets, we used Kruskal-Wallis $H$ tests on positions of the four RT for the two data sets, which resulted in significant differences.
	We conducted Mann-Whitney $U$ tests on the positions of $R_0$ and each of the three baselines resulting from all configurations together on the respective data sets. In the BTW'17 data set, $R_0$ performed significantly better than $B_{t3}$ but no significant differences were found when compared to the two other baselines.
	In the ECIR'17 data set, $R_0$ performed significantly better than all three baselines.
	
	\begin{table*}[t]
		\caption{Average positions (pos) of sets computed by the different configurations (conf) in the qualitative evaluation as well as the average number of relevant reviewers (\#rel) and the average position of entries from the different RTs per set.}
		\label{tab:best_comb_s1}
		\centering
		\begin{tabular}{l"l|l"l|l"l|l"l|l"l|l"l|l"l|l"l|l}
			data set  &  \multicolumn{8}{c"}{\textbf{BTW'17}} &  \multicolumn{8}{c}{\textbf{ECIR'17}}\\ 
			result type & \multicolumn{2}{c"}{$R_0$} & \multicolumn{2}{c"}{$B_{t3}$} & \multicolumn{2}{c"}{$B_{tr}$} & \multicolumn{2}{c"}{$B_{r}$} & \multicolumn{2}{c"}{$R_0$} & \multicolumn{2}{c"}{$B_{t3}$} & \multicolumn{2}{c"}{$B_{tr}$} & \multicolumn{2}{c}{$B_{r}$}\\ \hline
			avg. position $\forall$ conf& \multicolumn{2}{l"}{\textbf{
					1.5833}} & \multicolumn{2}{l"}{1.95}& \multicolumn{2}{l"}{2.933} &\multicolumn{2}{l"}{3.5167} & \multicolumn{2}{l"}{\textbf{1.2667}} &  \multicolumn{2}{l"}{1.3333} & \multicolumn{2}{l"}{1.6167} & \multicolumn{2}{l}{3.2167}\\ \hline
			conf$\backslash$measure  & \#rel & pos &\#rel & pos & \#rel & pos & \#rel & pos  & \#rel & pos &\#rel & pos & \#rel &  pos & \#rel & pos\\ 
			\hline 
			
			\hline
			
			tf-idf + $SUM$ ($b_1$) & .9& 1.7&1.45& 1.95& .95& 3.15& .8& 3.2  & 1.6 & 1.6& 1.55& 1.45& 1.25& 1.55&.4& 3.1\\
			tf-idf + $MNZ$ ($e_1$) &.85 & 1.55&1.4& 1.85& 1.05& 3.0& .6& 3.6  & 1.55 & \textbf{1.1}& 2.1& 1.15& 1.3& 1.4&.5& 3.15\\
			DBOW + $Votes_{\delta=.5}$ ($e_2$) & .8& \textbf{1.5}&1.25& 2.05&.75& 2.65& .6& 3.75 & 1.6 & \textbf{1.1}&2.05& 1.4& 1.3& 1.9&.9& 3.4\\
			
		\end{tabular}
	\end{table*}{}
	
	In a second manual evaluation of Step 2, we examined the best combinations from Step 1 ($b_1$, $e_1$ and $e_2$) with CT = full, k = 20, t = .5 as the best performing combinations from Step 2 performed bad in Step 1. Table~\ref{tab:best_comb_s1} was constructed exactly as described previously for Table~\ref{tab:mane2}.
	For both data sets, the best performing RT is $R_0$. It achieves the best average position for all configurations together and the combination resulting in the best position is $e_2$ with $R_0$. For ECIR'17, $R_0$ also produces the best overall position for $e_1$.

	\begin{table}[t]
		\caption{Configuration (c) and RTs with corresponding scores per data set and manually assessed average values $\in [0, 1]$ 
			(with 1 being the best possible and 0 being the worst possible value) 
			for aspects of sets for the twenty evaluated papers. $mA$: 1/3 $\forall$ reviewers with $h$ index $\ge$ 25; $mS$: each 1/3 if set contains at least one senior researcher, at least one junior researcher or at least one mid-career researcher; $mI$: 1/3 $\forall$ reviewers who published a relevant paper in the seven previous years; $mD$: 1/3 $\forall$ reviewer pairs without overlap in their work; $mE$: 1/3 for each relevant reviewer in the set. 
			Value $mSC$ is calculated similarly to SC, all manually evaluated aspects are multiplied.}
		\label{tab:step1_val_step2}
		\centering
		\begin{tabular}{l"l|l|l|l|l|l|l}
			&  \multicolumn{7}{c}{\textbf{BTW'17}}\\ 
			c$\times$RT & SC & $mSC$ & $mA$ & $mS$ & $mI$ & $mD$ & $mE$\\ 
			\hline 
			
			\hline
			
			$b_1$$\times$$R_0$ & \textbf{.0316} & .0383 & .8667& .6667& .25& .8833 & .3 \\  
			$b_1$$\times$$B_{t3}$ & .0089 & \textbf{.0767} & .8333& .6333& .3167& .95 & .4833\\   
			$b_1$$\times$$B_{tr}$ & .0079 & .0412&.7333& .7333& .25& .9667 & .3167\\   
			$b_1$$\times$$B_{r}$ & .0056 & .024& .65& .6833& .2167& .9333 & .2667\\ \hline 
			
			$e_1$$\times$$R_0$ & \textbf{.0308} & .0397 & .9833& .6667& .2333& .9167 & .2833\\  
			$e_1$$\times$$B_{t3}$ & .0107 & \textbf{.0667} & .9833& .55& .2833& .9333 & .4667\\   
			$e_1$$\times$$B_{tr}$ & .0082 & .0446 & .8333& .6667& .2333& .9833 & .35\\   
			$e_1$$\times$$B_{r}$ & .0085 & .0142 & .6667& .7333& .15& .9667 & .2\\ \hline
			
			$e_2$$\times$$R_0$ & \textbf{.0296} & .0338& 1& .6333& .2& 1 & .2667\\  
			$e_2$$\times$$B_{t3}$ & .0078 & \textbf{.0704}& .9& .65& .3333& .8667 & .4167\\   
			$e_2$$\times$$B_{tr}$ & .0082 & .0223 & .8167& .6667& .1667& .9833 & .25\\   
			$e_2$$\times$$B_{r}$ & .0069 & .0076 & .45& .75& .1167& .9667 & .2\\  
			\hline
			
			\hline
			& \multicolumn{7}{c}{\textbf{ECIR'17}}\\ 
			c$\times$RT & $SC$ & $mSC$ & $mA$ & $mS$ & $mI$ & $mD$ & $mE$\\ 
			\hline 
			
			\hline
			
			$b_1$$\times$$R_0$ & \textbf{.0337} & \textbf{.1539}& .8667& .6667& .5167& .9667 & .5333\\  
			$b_1$$\times$$B_{t3}$ & .0087 & .1182 & .8667& .55& .5333& .9& .5167\\   
			$b_1$$\times$$B_{tr}$ & .009 & .0753& .85& .6& .3667& .9667 &.4167\\   
			$b_1$$\times$$B_{r}$ & .0072 & .006 & .55& .6167& .1333& 1 & .1333\\ \hline 
			
			$e_1$$\times$$R_0$ & \textbf{.03} & .122 & 1& .5667& .4166& 1& .5167 \\  
			$e_1$$\times$$B_{t3}$ & .0082 & \textbf{.2432} &  1& .6167& .65& .8667& .7\\   
			$e_1$$\times$$B_{tr}$ & .0087 & .1097 & .9667& .65& .4167& .9667&.4333\\   
			$e_1$$\times$$B_{r}$ & .0043 & .0089&  .6667& .6833& .1167& 1& .1667\\ \hline
			
			$e_2$$\times$$R_0$ & \textbf{.0271} & .1493 & 1& .6& .4667& 1& .5333\\  
			$e_2$$\times$$B_{t3}$ & .0096 & \textbf{.1686} & .9167& .5& .5667& .95& .6833\\   
			$e_2$$\times$$B_{tr}$ & .007 & .084 & .8667& .65& .35& .9833& .4333\\   
			$e_2$$\times$$B_{r}$ & .0044 & .0297 &  .5667& .6667& .2667& .9833& .3\\  
			
		\end{tabular}
	\end{table}{}
	
	To better understand the impact of the five aspects, a human assessor also evaluated the quality of the results with the best combinations from Step 1 with respect to each aspect, assigning a value between 0 and 1 for each aspect. Table~\ref{tab:step1_val_step2} depicts average scores according to Equation~\ref{eq1} for combinations of the three best methods from Step 1 with all result types, manually assessed average values for the five aspects and ``manual'' scores computed by multiplying the per-aspect values. 
	In this evaluation, we wanted to compare the manually constructed scores to the automatic ones and evaluate possible effects of opposing aspects. We observe vast differences in the manual scores $mSC$ and the computed $scores$,
	in almost all cases $B_{t3}$ achieves the highest $mSC$. As we have already seen in Tables~\ref{tab:mane2} and ~\ref{tab:best_comb_s1}, $R_0$ generally achieves the best average positions for sets of reviewers. This discrepancy further underlines the suitability of our approach. RevASIDE produces reviewer sets based on calculated aspects which are preferable in a manual evaluation to the sets from $B_{t3}$ which achieved the highest $mSC$ in the manual assessment of aspects.
	
	We found a positive correlation of aspects $mA$ and $mI$ (.597 for BTW'17, .802 for ECIR'17) which is significant with Pearson's correlation coefficient for both data sets. A higher authority might be equivalent to a higher number of papers, especially in the last seven years which might increase the probability of one of these papers being from the area of the manuscript and thus signals reviewers' interest.
	Also for both data sets, the negative correlation between $mI$ and $mD$ (-.589 for BTW'17, -.72 for ECIR'17) is significant with Pearson's correlation coefficient. If reviewers in a set are very interested in a manuscript, it seems likely that the set is not as diverse. In BTW'17, $mA$ is significantly correlated with $mS$ (-.789), in ECIR'17 this negative correlation is not significant with Pearson's correlation coefficient. This observation can be explained as sets having high authority normally consist solely of researchers with high seniority. 
	We found opposing objectives coded into the aspects which might have lead to methods from Table~\ref{tab:best_comb} achieving low results in Step 1 but being useful in Step 2.
	
	In general average positions of sets from the different RTs are highly dependent on the configuration in BTW'17 and ECIR'17 for the best performing configurations in Step 2 but the overall best results are achieved independent of configuration by $R_0$. 
	From these observations we conclude that $R_0$ and thereby RevASIDE is a well-performing solution of the reviewer set assignment problem which is generally applicable. Thus, hypothesis $H_6$ is verified.
	
	\section{Conclusion and Future Work}
	\label{con}
	
	In this paper we proposed and evaluated RevASIDE, a method for assigning complementing reviewer sets for submissions from fixed candidate pools. Our approach incorporates authority, seniority, interests of researchers, diversity of the reviewer set as well as candidates' expertise. Additionally, we presented three new data sets suitable for reviewer set recommendation. 
	
	In this context we examine the expert search as well as the reviewer set assignment tasks and show RevASIDE's general applicability: for the first task we revaluated expert voting techniques utilising different document representations. We verified the general usefulness of Step 1 for the expert search (addressed with hypothesis $H_1$) and reviewer set recommendation task (addressed with hypothesis $H_5$). Additionally, we have shown the suitability of simple textual similarity methods utilising tf-idf compared to more advanced techniques using BERT, which in terms rejected hypothesis $H_2$.
	For the second task RevASIDE produces significantly higher overall scores for reviewer set assignment compared to three baselines in an quantitative evaluation which shows the approach's usefulness. 
	In a qualitative evaluation we observed that sets assembled by our system are generally significantly more suitable recommendations compared to our three baselines. 
	We were able to confirm the results from the quantitative evaluation and thus verified $H_6$.
	
	Possible extensions might include weighting the different quantifiable aspects defined in Step 2 of the approach and incorporating the venue which reviewers are recommended for. The number of assigned reviewers could be varied for each submission to take into account papers with broad content.
	
	Future work will focus on recommending suitable reviewer sets for whole venues. Here, the optimisation problem of single manus\-cripts is extended to include all manuscripts and several constraints such as individually differing maximal numbers of papers per reviewer come into consideration. Such an approach should also consider fairness~\cite{fair} of the recommended reviewer sets.
	It would be interesting to observe gaps in the expertise displayed by the program committee in terms of fit with submitted manuscripts together with suggesting new reviewers matching the missing criteria. 
	Another feasible extension might be the recommendation of a program committee based on former and recent conferences and anticipated submissions. Here, topical development between years is important. 
	Furthermore, explainability~\cite{DBLP:journals/ftir/ZhangC20} of the recommended reviewer sets should be a priority. In our case, radar charts could be used for example to visualise the values which the sets achieved in the different quantifiable aspects.

	\bibliographystyle{ACM-Reference-Format}

\begin{thebibliography}{26}
	
	
	\ifx \showCODEN    \undefined \def \showCODEN     #1{\unskip}     \fi
	\ifx \showDOI      \undefined \def \showDOI       #1{#1}\fi
	\ifx \showISBNx    \undefined \def \showISBNx     #1{\unskip}     \fi
	\ifx \showISBNxiii \undefined \def \showISBNxiii  #1{\unskip}     \fi
	\ifx \showISSN     \undefined \def \showISSN      #1{\unskip}     \fi
	\ifx \showLCCN     \undefined \def \showLCCN      #1{\unskip}     \fi
	\ifx \shownote     \undefined \def \shownote      #1{#1}          \fi
	\ifx \showarticletitle \undefined \def \showarticletitle #1{#1}   \fi
	\ifx \showURL      \undefined \def \showURL       {\relax}        \fi
	\providecommand\bibfield[2]{#2}
	\providecommand\bibinfo[2]{#2}
	\providecommand\natexlab[1]{#1}
	\providecommand\showeprint[2][]{arXiv:#2}
	
	\bibitem[\protect\citeauthoryear{Ailamaki, Chrysogelos, Deshpande, and
		Kraska}{Ailamaki et~al\mbox{.}}{2019}]%
	{sigmod}
	\bibfield{author}{\bibinfo{person}{Anastasia Ailamaki},
		\bibinfo{person}{Periklis Chrysogelos}, \bibinfo{person}{Amol Deshpande},
		{and} \bibinfo{person}{Tim Kraska}.} \bibinfo{year}{2019}\natexlab{}.
	\newblock \showarticletitle{The {SIGMOD} 2019 Research Track Reviewing System}.
	\newblock \bibinfo{journal}{\emph{{SIGMOD} Record}} \bibinfo{volume}{48},
	\bibinfo{number}{2} (\bibinfo{year}{2019}), \bibinfo{pages}{47--54}.
	\newblock
	
	
	\bibitem[\protect\citeauthoryear{Blei, Ng, and Jordan}{Blei
		et~al\mbox{.}}{2003}]%
	{lda}
	\bibfield{author}{\bibinfo{person}{David~M. Blei}, \bibinfo{person}{Andrew~Y.
			Ng}, {and} \bibinfo{person}{Michael~I. Jordan}.}
	\bibinfo{year}{2003}\natexlab{}.
	\newblock \showarticletitle{Latent Dirichlet Allocation}.
	\newblock \bibinfo{journal}{\emph{J. Mach. Learn. Res.}}  \bibinfo{volume}{3}
	(\bibinfo{year}{2003}), \bibinfo{pages}{993--1022}.
	\newblock
	
	
	\bibitem[\protect\citeauthoryear{Charlin and Zemel}{Charlin and Zemel}{2013}]%
	{toronto}
	\bibfield{author}{\bibinfo{person}{Laurent Charlin} {and}
		\bibinfo{person}{Richard~S. Zemel}.} \bibinfo{year}{2013}\natexlab{}.
	\newblock \showarticletitle{The Toronto Paper Matching System: An automated
		paper-reviewer assignment system}.
	\newblock
	
	
	\bibitem[\protect\citeauthoryear{Chughtai, Lee, Shahzadi, Kabir, and
		Hassan}{Chughtai et~al\mbox{.}}{2020}]%
	{chughtai}
	\bibfield{author}{\bibinfo{person}{Gohar~Rehman Chughtai}, \bibinfo{person}{Jia
			Lee}, \bibinfo{person}{Mahnoor Shahzadi}, \bibinfo{person}{Asif Kabir}, {and}
		\bibinfo{person}{Muhammad Arshad~Shehzad Hassan}.}
	\bibinfo{year}{2020}\natexlab{}.
	\newblock \showarticletitle{An efficient ontology-based topic-specific article
		recommendation model for best-fit reviewers}.
	\newblock \bibinfo{journal}{\emph{Scientometrics}} \bibinfo{volume}{122},
	\bibinfo{number}{1} (\bibinfo{year}{2020}), \bibinfo{pages}{249--265}.
	\newblock
	
	
	\bibitem[\protect\citeauthoryear{Devlin, Chang, Lee, and Toutanova}{Devlin
		et~al\mbox{.}}{2019}]%
	{BERT}
	\bibfield{author}{\bibinfo{person}{Jacob Devlin}, \bibinfo{person}{Ming{-}Wei
			Chang}, \bibinfo{person}{Kenton Lee}, {and} \bibinfo{person}{Kristina
			Toutanova}.} \bibinfo{year}{2019}\natexlab{}.
	\newblock \showarticletitle{{BERT:} Pre-training of Deep Bidirectional
		Transformers for Language Understanding}. In
	\bibinfo{booktitle}{\emph{{NAACL-HLT} 2019}}. \bibinfo{publisher}{ACL},
	\bibinfo{pages}{4171--4186}.
	\newblock
	
	
	\bibitem[\protect\citeauthoryear{Ishag, Park, Lee, and Ryu}{Ishag
		et~al\mbox{.}}{2019}]%
	{ishag}
	\bibfield{author}{\bibinfo{person}{Musa Ibrahim~M. Ishag},
		\bibinfo{person}{Kwang{-}Ho Park}, \bibinfo{person}{Jong~Yun Lee}, {and}
		\bibinfo{person}{Keun~Ho Ryu}.} \bibinfo{year}{2019}\natexlab{}.
	\newblock \showarticletitle{A Pattern-Based Academic Reviewer Recommendation
		Combining Author-Paper and Diversity Metrics}.
	\newblock \bibinfo{journal}{\emph{{IEEE} Access}}  \bibinfo{volume}{7}
	(\bibinfo{year}{2019}), \bibinfo{pages}{16460--16475}.
	\newblock
	
	
	\bibitem[\protect\citeauthoryear{J{\"{a}}rvelin and
		Kek{\"{a}}l{\"{a}}inen}{J{\"{a}}rvelin and Kek{\"{a}}l{\"{a}}inen}{2002}]%
	{DBLP:journals/tois/JarvelinK02}
	\bibfield{author}{\bibinfo{person}{Kalervo J{\"{a}}rvelin} {and}
		\bibinfo{person}{Jaana Kek{\"{a}}l{\"{a}}inen}.}
	\bibinfo{year}{2002}\natexlab{}.
	\newblock \showarticletitle{Cumulated gain-based evaluation of {IR}
		techniques}.
	\newblock \bibinfo{journal}{\emph{{ACM} Trans. Inf. Syst.}}
	\bibinfo{volume}{20}, \bibinfo{number}{4} (\bibinfo{year}{2002}),
	\bibinfo{pages}{422--446}.
	\newblock
	
	
	\bibitem[\protect\citeauthoryear{Jin, Niu, Ji, and Geng}{Jin
		et~al\mbox{.}}{2020}]%
	{trendResInt2}
	\bibfield{author}{\bibinfo{person}{Jian Jin}, \bibinfo{person}{Baozhuang Niu},
		\bibinfo{person}{Ping Ji}, {and} \bibinfo{person}{Qian Geng}.}
	\bibinfo{year}{2020}\natexlab{}.
	\newblock \showarticletitle{An integer linear programming model of reviewer
		assignment with research interest considerations}.
	\newblock \bibinfo{journal}{\emph{Ann. Oper. Res.}} \bibinfo{volume}{291},
	\bibinfo{number}{1} (\bibinfo{year}{2020}), \bibinfo{pages}{409--433}.
	\newblock
	
	
	\bibitem[\protect\citeauthoryear{Kou, U, Mamoulis, Li, Li, and Gong}{Kou
		et~al\mbox{.}}{2015}]%
	{DBLP:journals/pvldb/KouUMLLG15}
	\bibfield{author}{\bibinfo{person}{Ngai~Meng Kou}, \bibinfo{person}{Leong~Hou
			U}, \bibinfo{person}{Nikos Mamoulis}, \bibinfo{person}{Yuhong Li},
		\bibinfo{person}{Ye Li}, {and} \bibinfo{person}{Zhiguo Gong}.}
	\bibinfo{year}{2015}\natexlab{}.
	\newblock \showarticletitle{A Topic-based Reviewer Assignment System}.
	\newblock \bibinfo{journal}{\emph{Proc. {VLDB} Endow.}} \bibinfo{volume}{8},
	\bibinfo{number}{12} (\bibinfo{year}{2015}), \bibinfo{pages}{1852--1855}.
	\newblock
	
	
	\bibitem[\protect\citeauthoryear{Kreutz, Sahitaj, and Schenkel}{Kreutz
		et~al\mbox{.}}{2020}]%
	{kreutz}
	\bibfield{author}{\bibinfo{person}{Christin~Katharina Kreutz},
		\bibinfo{person}{Premtim Sahitaj}, {and} \bibinfo{person}{Ralf Schenkel}.}
	\bibinfo{year}{2020}\natexlab{}.
	\newblock \showarticletitle{Evaluating semantometrics from computer science
		publications}.
	\newblock \bibinfo{journal}{\emph{Scientometrics}} \bibinfo{volume}{125},
	\bibinfo{number}{3} (\bibinfo{year}{2020}), \bibinfo{pages}{2915--2954}.
	\newblock
	
	
	\bibitem[\protect\citeauthoryear{Lau and Baldwin}{Lau and Baldwin}{2016}]%
	{numdim}
	\bibfield{author}{\bibinfo{person}{Jey~Han Lau} {and} \bibinfo{person}{Timothy
			Baldwin}.} \bibinfo{year}{2016}\natexlab{}.
	\newblock \showarticletitle{An Empirical Evaluation of doc2vec with Practical
		Insights into Document Embedding Generation}. In
	\bibinfo{booktitle}{\emph{Rep4NLP@ACL 2016}}. \bibinfo{publisher}{ACL},
	\bibinfo{pages}{78--86}.
	\newblock
	
	
	\bibitem[\protect\citeauthoryear{Le and Mikolov}{Le and Mikolov}{2014}]%
	{d2v}
	\bibfield{author}{\bibinfo{person}{Quoc~V. Le} {and} \bibinfo{person}{Tomas
			Mikolov}.} \bibinfo{year}{2014}\natexlab{}.
	\newblock \showarticletitle{Distributed Representations of Sentences and
		Documents}. In \bibinfo{booktitle}{\emph{{ICML} 2014}}
	\emph{(\bibinfo{series}{{JMLR} Workshop and Conference Proceedings})},
	Vol.~\bibinfo{volume}{32}. \bibinfo{publisher}{JMLR.org},
	\bibinfo{pages}{1188--1196}.
	\newblock
	
	
	\bibitem[\protect\citeauthoryear{Ley}{Ley}{2009}]%
	{ley}
	\bibfield{author}{\bibinfo{person}{Michael Ley}.}
	\bibinfo{year}{2009}\natexlab{}.
	\newblock \showarticletitle{{DBLP} - Some Lessons Learned}.
	\newblock \bibinfo{journal}{\emph{{PVLDB}}} \bibinfo{volume}{2},
	\bibinfo{number}{2} (\bibinfo{year}{2009}), \bibinfo{pages}{1493--1500}.
	\newblock
	
	
	\bibitem[\protect\citeauthoryear{Liu, Suel, and Memon}{Liu
		et~al\mbox{.}}{2014}]%
	{robustModel}
	\bibfield{author}{\bibinfo{person}{Xiang Liu}, \bibinfo{person}{Torsten Suel},
		{and} \bibinfo{person}{Nasir~D. Memon}.} \bibinfo{year}{2014}\natexlab{}.
	\newblock \showarticletitle{A robust model for paper reviewer assignment}. In
	\bibinfo{booktitle}{\emph{RecSys 2014}}. \bibinfo{publisher}{{ACM}},
	\bibinfo{pages}{25--32}.
	\newblock
	
	
	\bibitem[\protect\citeauthoryear{Macdonald and Ounis}{Macdonald and
		Ounis}{2006}]%
	{macdonald}
	\bibfield{author}{\bibinfo{person}{Craig Macdonald} {and} \bibinfo{person}{Iadh
			Ounis}.} \bibinfo{year}{2006}\natexlab{}.
	\newblock \showarticletitle{Voting for candidates: adapting data fusion
		techniques for an expert search task}. In \bibinfo{booktitle}{\emph{{CIKM}
			2006}}. \bibinfo{publisher}{{ACM}}, \bibinfo{pages}{387--396}.
	\newblock
	
	
	\bibitem[\protect\citeauthoryear{Machado and Stefanidis}{Machado and
		Stefanidis}{2019}]%
	{fair}
	\bibfield{author}{\bibinfo{person}{Lucas Machado} {and} \bibinfo{person}{Kostas
			Stefanidis}.} \bibinfo{year}{2019}\natexlab{}.
	\newblock \showarticletitle{Fair Team Recommendations for Multidisciplinary
		Projects}. In \bibinfo{booktitle}{\emph{{WI} 2019}}.
	\bibinfo{publisher}{{ACM}}, \bibinfo{pages}{293--297}.
	\newblock
	\showISBNx{978-1-4503-6934-3}
	
	
	\bibitem[\protect\citeauthoryear{Maleszka, Maleszka, Kr{\'{o}}l, Hernes,
		Martins, Homann, and Vossen}{Maleszka et~al\mbox{.}}{2020}]%
	{DBLP:conf/aciids/MaleszkaMKHMHV20}
	\bibfield{author}{\bibinfo{person}{Marcin Maleszka},
		\bibinfo{person}{Bernadetta Maleszka}, \bibinfo{person}{Dariusz Kr{\'{o}}l},
		\bibinfo{person}{Marcin Hernes}, \bibinfo{person}{Denis Mayr~Lima Martins},
		\bibinfo{person}{Leschek Homann}, {and} \bibinfo{person}{Gottfried Vossen}.}
	\bibinfo{year}{2020}\natexlab{}.
	\newblock \showarticletitle{A Modular Diversity Based Reviewer Recommendation
		System}. In \bibinfo{booktitle}{\emph{{ACIIDS} 2020}}
	\emph{(\bibinfo{series}{Communications in Computer and Information
			Science})}, Vol.~\bibinfo{volume}{1178}. \bibinfo{publisher}{Springer},
	\bibinfo{pages}{550--561}.
	\newblock
	\showISBNx{978-981-15-3379-2}
	
	
	\bibitem[\protect\citeauthoryear{Papagelis, Plexousakis, and
		Nikolaou}{Papagelis et~al\mbox{.}}{2005}]%
	{DBLP:conf/wise/PapagelisPN05}
	\bibfield{author}{\bibinfo{person}{Manos Papagelis}, \bibinfo{person}{Dimitris
			Plexousakis}, {and} \bibinfo{person}{Panagiotis Nikolaou}.}
	\bibinfo{year}{2005}\natexlab{}.
	\newblock \showarticletitle{{CONFIOUS:} Managing the Electronic Submission and
		Reviewing Process of Scientific Conferences}. In
	\bibinfo{booktitle}{\emph{{WISE} 2005}} \emph{(\bibinfo{series}{LNCS})},
	Vol.~\bibinfo{volume}{3806}. \bibinfo{publisher}{Springer},
	\bibinfo{pages}{711--720}.
	\newblock
	\showISBNx{3-540-30017-1}
	
	
	\bibitem[\protect\citeauthoryear{Sakr, Ragab, Maher, and Awad}{Sakr
		et~al\mbox{.}}{2019}]%
	{minaret}
	\bibfield{author}{\bibinfo{person}{Sherif Sakr}, \bibinfo{person}{Mohamed
			Ragab}, \bibinfo{person}{Mohamed Maher}, {and} \bibinfo{person}{Ahmed Awad}.}
	\bibinfo{year}{2019}\natexlab{}.
	\newblock \showarticletitle{{MINARET:} {A} Recommendation Framework for
		Scientific Reviewers}. In \bibinfo{booktitle}{\emph{{EDBT} 2019}}.
	\bibinfo{publisher}{OpenProceedings.org}, \bibinfo{pages}{538--541}.
	\newblock
	
	
	\bibitem[\protect\citeauthoryear{Sinha, Shen, Song, Ma, Eide, Hsu, and
		Wang}{Sinha et~al\mbox{.}}{2015}]%
	{DBLP:conf/www/SinhaSSMEHW15}
	\bibfield{author}{\bibinfo{person}{Arnab Sinha}, \bibinfo{person}{Zhihong
			Shen}, \bibinfo{person}{Yang Song}, \bibinfo{person}{Hao Ma},
		\bibinfo{person}{Darrin Eide}, \bibinfo{person}{Bo{-}June~Paul Hsu}, {and}
		\bibinfo{person}{Kuansan Wang}.} \bibinfo{year}{2015}\natexlab{}.
	\newblock \showarticletitle{An Overview of Microsoft Academic Service {(MAS)}
		and Applications}. In \bibinfo{booktitle}{\emph{{WWW} (Companion Volume)}}.
	\bibinfo{publisher}{{ACM}}, \bibinfo{pages}{243--246}.
	\newblock
	
	
	\bibitem[\protect\citeauthoryear{Tang, Zhang, Yao, Li, Zhang, and Su}{Tang
		et~al\mbox{.}}{2008}]%
	{aminer}
	\bibfield{author}{\bibinfo{person}{Jie Tang}, \bibinfo{person}{Jing Zhang},
		\bibinfo{person}{Limin Yao}, \bibinfo{person}{Juanzi Li}, \bibinfo{person}{Li
			Zhang}, {and} \bibinfo{person}{Zhong Su}.} \bibinfo{year}{2008}\natexlab{}.
	\newblock \showarticletitle{ArnetMiner: extraction and mining of academic
		social networks}. In \bibinfo{booktitle}{\emph{{SIGKDD} 2008}}.
	\bibinfo{publisher}{{ACM}}, \bibinfo{pages}{990--998}.
	\newblock
	
	
	\bibitem[\protect\citeauthoryear{Yang, Liu, Yi, Chen, and Niu}{Yang
		et~al\mbox{.}}{2020}]%
	{DBLP:journals/asc/YangLYCN20}
	\bibfield{author}{\bibinfo{person}{Chen Yang}, \bibinfo{person}{Tingting Liu},
		\bibinfo{person}{Wenjie Yi}, \bibinfo{person}{Xiaohong Chen}, {and}
		\bibinfo{person}{Ben Niu}.} \bibinfo{year}{2020}\natexlab{}.
	\newblock \showarticletitle{Identifying expertise through semantic modeling:
		{A} modified {BBPSO} algorithm for the reviewer assignment problem}.
	\newblock \bibinfo{journal}{\emph{Appl. Soft Comput.}}  \bibinfo{volume}{94}
	(\bibinfo{year}{2020}), \bibinfo{pages}{106483}.
	\newblock
	
	
	\bibitem[\protect\citeauthoryear{Yang, Kuo, Lee, and Ho}{Yang
		et~al\mbox{.}}{2009}]%
	{colabIntell}
	\bibfield{author}{\bibinfo{person}{Kai{-}Hsiang Yang},
		\bibinfo{person}{Tai{-}Liang Kuo}, \bibinfo{person}{Hahn{-}Ming Lee}, {and}
		\bibinfo{person}{Jan{-}Ming Ho}.} \bibinfo{year}{2009}\natexlab{}.
	\newblock \showarticletitle{A Reviewer Recommendation System Based on
		Collaborative Intelligence}. In \bibinfo{booktitle}{\emph{{WI} 2009}}.
	\bibinfo{publisher}{{IEEE} Computer Society}, \bibinfo{pages}{564--567}.
	\newblock
	
	
	\bibitem[\protect\citeauthoryear{Zhang, Zhao, Duan, Chen, Zhang, and
		Tang}{Zhang et~al\mbox{.}}{2020}]%
	{mlClass}
	\bibfield{author}{\bibinfo{person}{Dong Zhang}, \bibinfo{person}{Shu Zhao},
		\bibinfo{person}{Zhen Duan}, \bibinfo{person}{Jie Chen},
		\bibinfo{person}{Yanping Zhang}, {and} \bibinfo{person}{Jie Tang}.}
	\bibinfo{year}{2020}\natexlab{}.
	\newblock \showarticletitle{A Multi-Label Classification Method Using a
		Hierarchical and Transparent Representation for Paper-Reviewer
		Recommendation}.
	\newblock \bibinfo{journal}{\emph{TOIS}}  \bibinfo{volume}{38}
	(\bibinfo{date}{02} \bibinfo{year}{2020}), \bibinfo{pages}{1--20}.
	\newblock
	
	
	\bibitem[\protect\citeauthoryear{Zhang and Chen}{Zhang and Chen}{2020}]%
	{DBLP:journals/ftir/ZhangC20}
	\bibfield{author}{\bibinfo{person}{Yongfeng Zhang} {and} \bibinfo{person}{Xu
			Chen}.} \bibinfo{year}{2020}\natexlab{}.
	\newblock \showarticletitle{Explainable Recommendation: {A} Survey and New
		Perspectives}.
	\newblock \bibinfo{journal}{\emph{Found. Trends Inf. Retr.}}
	\bibinfo{volume}{14}, \bibinfo{number}{1} (\bibinfo{year}{2020}),
	\bibinfo{pages}{1--101}.
	\newblock
	
	
	\bibitem[\protect\citeauthoryear{Zhao, Zhang, Duan, Chen, Zhang, and Tang}{Zhao
		et~al\mbox{.}}{2018}]%
	{Zhao}
	\bibfield{author}{\bibinfo{person}{Shu Zhao}, \bibinfo{person}{Dong Zhang},
		\bibinfo{person}{Zhen Duan}, \bibinfo{person}{Jie Chen},
		\bibinfo{person}{Yan{-}Ping Zhang}, {and} \bibinfo{person}{Jie Tang}.}
	\bibinfo{year}{2018}\natexlab{}.
	\newblock \showarticletitle{A novel classification method for paper-reviewer
		recommendation}.
	\newblock \bibinfo{journal}{\emph{Scientometrics}} \bibinfo{volume}{115},
	\bibinfo{number}{3} (\bibinfo{year}{2018}), \bibinfo{pages}{1293--1313}.
	\newblock
	
	
\end{thebibliography}

\end{document}